\documentclass[journal]{IEEEtran}
\usepackage{epstopdf}% To incorporate .eps illustrations using PDFLaTeX, etc.
\usepackage{subfigure}% Support for small, `sub' figures and tables
\usepackage{multirow}
\usepackage{amsfonts}
\usepackage{mathrsfs}
\usepackage{amssymb}
\usepackage{epstopdf}% To incorporate .eps illustrations using PDFLaTeX, etc.
\usepackage{subfigure}% Support for small, `sub' figures and tables
\usepackage[table]{xcolor}
\usepackage{color}
\usepackage{colortbl}
\usepackage{tabularx}
\usepackage{bm}
\usepackage{booktabs}
\usepackage{cite}
\ifCLASSINFOpdf
 \usepackage[pdftex]{graphicx}
\else
\fi
\begin{document}
\title{Cross-Domain Collaborative Learning via Cluster Canonical Correlation Analysis and Random Walker for Hyperspectral Image Classification}
\author{Yao~Qin,~\IEEEmembership{Student Member,~IEEE},
       ~Lorenzo~Bruzzone,~\IEEEmembership{Fellow,~IEEE},
      ~Biao~Li,
       and~Yuanxin~Ye,~\IEEEmembership{Member,~IEEE}
\thanks{Manuscript received Oct. 30, 2018. (\emph{Corresponding author: Lorenzo Bruzzone}.)}
\thanks{Y. Qin is with the College of Electronic Science, National University of Defense Technology, Changsha 410073, China and Department of Information Engineering and Computer Science, University of Trento, 38123 Trento, Italy (e-mail: yao.qin@unitn.it).}
\thanks{L. Bruzzone is with Department of Information Engineering and Computer Science, University of Trento, 38123 Trento, Italy (e-mail: lorenzo.bruzzone@ing.unitn.it).}% <-this % stops a space
\thanks{B. Li is with the College of Electronic Science, National University of Defense Technology, Changsha 410073, China.}
\thanks{Y. Ye is with the Faculty of Geosciences and Environmental Engineering, Southwest Jiaotong University, Chengdu 610031, China (e-mail: yeyuanxin@home.swjtu.edu.cn).}% <-this % stops a space
}


\markboth{Journal of \LaTeX\ Class Files,~Vol.~14, No.~8, August~2015}%
{Shell \MakeLowercase{\textit{et al.}}: Bare Demo of IEEEtran.cls for IEEE Journals}



\maketitle
\begin{abstract}
This paper introduces a novel heterogenous domain adaptation (HDA) method for hyperspectral image classification with a limited amount of labeled samples in both domains. The method is achieved in the way of cross-domain collaborative learning (CDCL), which is addressed via cluster canonical correlation analysis (C-CCA) and random walker (RW) algorithms. To be specific, the proposed \textbf{CDCL} method is an iterative process of three main components, i.e. RW-based pseudolabeling, cross domain learning via C-CCA and final classification based on extended RW (ERW) algorithm. Firstly, given the initially labeled target samples as training set ($\mathbf{TS}$), the RW-based pseudolabeling is employed to update $\mathbf{TS}$ and extract target clusters ($\mathbf{TCs}$) by fusing the segmentation results obtained by RW and extended RW (ERW) classifiers. Secondly, cross domain learning via C-CCA is applied using labeled source samples and $\mathbf{TCs}$. The unlabeled target samples are then classified  with the estimated probability maps using the model trained in the projected correlation subspace. The newly estimated probability map and $\mathbf{TS}$ are used for updating $\mathbf{TS}$ again via RW-based pseudolabeling. Finally, when the iterative process converges, the result obtained by the ERW classifier using the final $\mathbf{TS}$ and estimated probability maps is regarded as the final classification map. Experimental results on four real HSIs demonstrate that the proposed method can achieve better performance compared with the state-of-the-art HDA and ERW methods.
\end{abstract}


% \fi
\IEEEpeerreviewmaketitle

\section{Introduction}
\IEEEPARstart{H}{perspectral} images (HSIs) can capture detailed spectral information measured in contiguous bands of the electromagnetic spectrum \cite{Fauvel2013,Camps-Valls2014,Tuia2016a} and have been widely used in various remote sensing applications, such as environmental monitoring \cite{Schneider2014} and mineral exploration \cite{Tiwari2011}. One fundamental challenge in these applications is to assign a unique label to each pixel in the image, which is called HSI classification. When the problem is treated as a supervised learning and solved using machine learning methods (including random forest \cite{Ham2005}, support vector machine (SVM) \cite{Melgani2004}, laplacian SVM (LapSVM) \cite{Belkin2006,Melacci2011,Yang2015}, decision trees \cite{Delalieux2012} and support tensor machine (STM) \cite{Guo2016}), a large amount of labeled samples are required due to the high dimensionality of hyperspectral data. This would require extensive and expensive field data collection campaigns. Consequently, only a small quantity of labeled samples are available in most practical applications of HSI classification. In order to solve the problem, several machine learning and feature extraction methods have been widely applied to hyperspectral data \cite{Camps-Valls2014}, such as active learning (AL) \cite{Rajan2008,Persello2012,Persello2014,Sun2015,Patra2017}, semi-supervised learning (SSL) \cite{Munoz-Mari2012,Dopido2013,Persello2014}, spectral-spatial classification \cite{Bruzzone2009,Bernard2012,Li2012,Zhong2015a}, domain adaptation (DA) \cite{Bruzzone2010,Matasci2015,Tuia2016a} and more recently deep learning based techniques \cite{Chen2016,Hao2018,Hao2018a}.
In this paper, we focus on applying DA to HSI classification.

According to the machine learning and pattern recognition literature, DA refers to solving the problem of adapting model trained on the source domain to the target domain. When applied to HSI classification, DA aims to generate accurate classification map of target HSI by utilizing the knowledge learned on the source HSI. According to \cite{Bruzzone2010}, unsupervised DA refers to the case where there are no labeled samples available in the target domain, whereas semi-supervised DA represents the case where there are few target labeled samples. Further, \textbf{h}eterogenous \textbf{DA} (HDA) refers to the dimensions of features in both domains are assumed to be different. Since we assume that a limited amount of labeled samples are available in the target HSI, we therefore focus on semi-supervised HDA for HSI classification. Although there are several HDA methods based on deep learning for visual and remote sensing applications \cite{Long2015,Rozantsev2018,Zhou2018}, 
the feature representation ability of deep learning models is strongly dependent on the availability of a large number of training samples \cite{Zhang2016b,Pan2017}.
Therefore, it is difficult to obtain a reliable deep learning model with the availability of very few samples in hyperdimensional feature spaces. Given the assumption of the limited number of training samples, in our paper we focus on handcrafted features for HDA.

In the literature of HDA, one of the simplest feature-based approaches is the feature augmentation proposed in \cite{DaumeIII2009}, which extended versions, called heterogeneous feature augmentation (HFA) and semi-supervised HFA (SHFA), have been recently proposed in \cite{Li2014}. 
In \cite{Jhuo2012}, a robust domain adaptation low-rank reconstruction method is introduced, where a transformed intermediate representation of the samples in the source domain is linearly reconstructed by the target samples.
In \cite{Nielsen2007}, the authors align domains with canonical correlation analysis (CCA) and then perform change detection. The approach is extended to a kernel and semisupervised version in \cite{Volpi2015}, where the authors perform change detection with different sensors. 
 In \cite{Samat2017}, the supervised multi-view canonical correlation analysis ensemble  is presented to address HDA problems.
In \cite{HubertTsai2016}, the proposed cross-domain landmark selection (CDLS) method is able to learn representative cross-domain landmarks for deriving a proper feature subspace for adaptation and classification purposes.
Different from the above feature-based category, several studies employ manifold learning to preserve the original geometry. 
In \cite{Wang2011}, the method of domain adaptation using manifold alignment (DAMA) can reuse labeled data from multiple source domains in the target domain even in the case when the input domains do not share any common features or instances.
In \cite{Tuia2014}, semi-supervised manifold alignment (SSMA) is proposed, where both domains are matched through manifold alignment while preserving label (dis)similarities and the geometric structures of the single manifold in both domains. Recently, the kernelized manifold alignment (KEMA) has been introduced in \cite{Tuia2016}. In \cite{Zhou2018}, a deep feature alignment neural network is introduced to carry out the domain adaptation, where discriminative features for the source and target domains are extracted using deep convolutional recurrent neural networks and then aligned with each other layer-by-layer. In \cite{Persello2016}, a kernel-based domain-invariant feature selection method has been proposed for the classification of hyperspectral images, where a novel measure of data shift for evaluating the domain stability is defined.

As stated earlier, it is not feasible to obtain a large amount of labeled target samples in practical applications. On the other hand, if a sufficient number of labeled samples are available in the target HSI, an accurate classification map can be achieved by using newly-developed deep learning methods \cite{Chen2015}. 
Therefore, it is reasonable to assume that only limited labeled samples can be used in the semi-supervised HDA problem. In oder to address the problem and obtain better classification performance, two key problems should be solved, i.e. how to obtain more pseudo-labeled reliable target samples for adaptation and how to achieve better adaptation with these samples.
 
\begin{figure}[!t]
\setlength{\abovecaptionskip}{-5pt}
\setlength{\belowcaptionskip}{-20pt}
\centering
\includegraphics[width=0.45\textwidth]{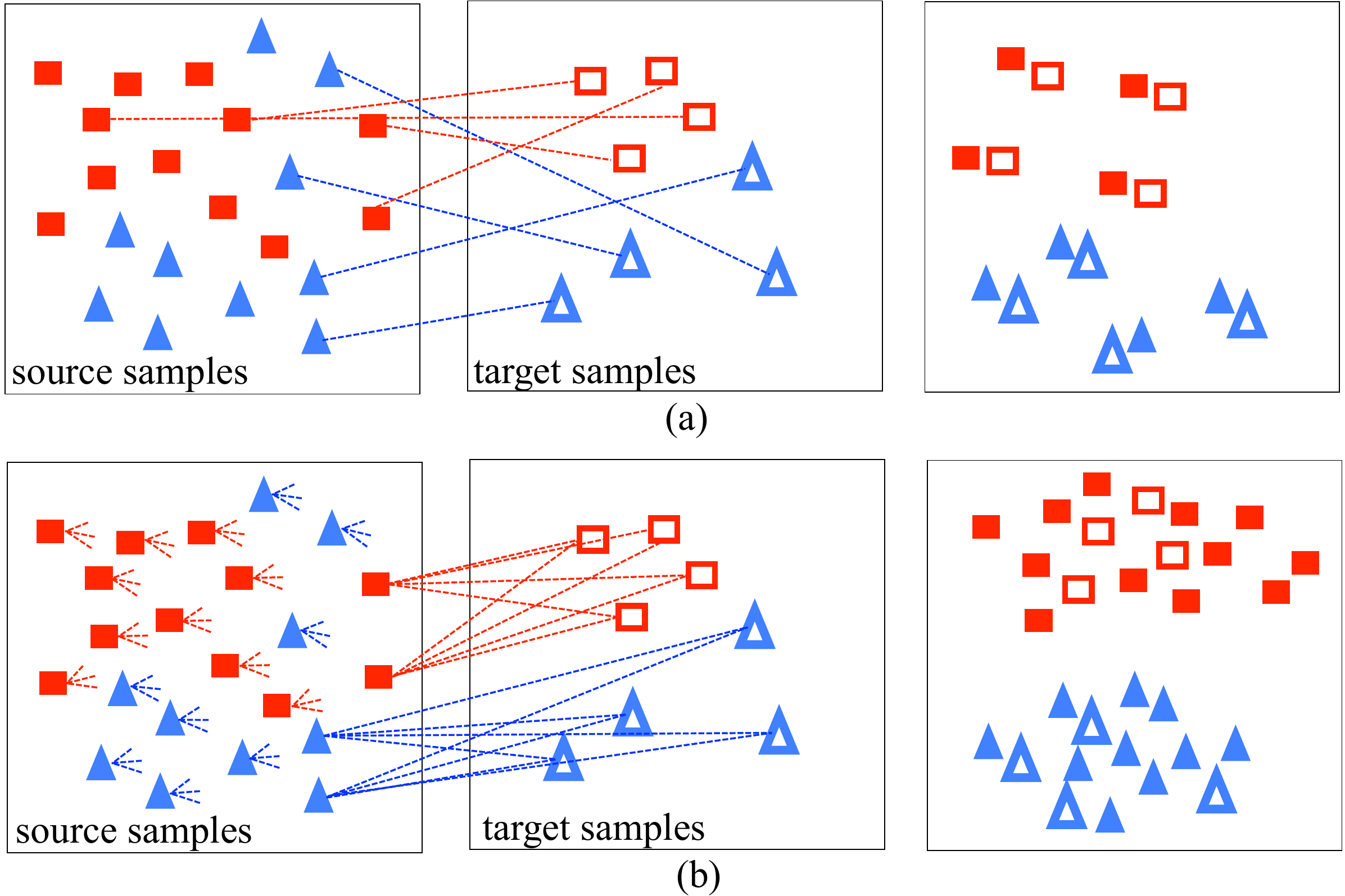}
\caption{Representation of CCA and C-CCA methods to obtain correlated subspaces between source and target samples. (a) CCA uses pairwise correspondences between source and target samples and can hardly segregate the two clusters. (b) C-CCA uses all pairwise correspondences within a cluster across the two sets of samples and results in cluster segregation.}
\label{CCCA}
\end{figure}

 In this paper, random walker (RW)-based pseudolabeling \cite{Sun2017} and cluster canonical correlation analysis (C-CCA) \cite{Rasiwasia2014} are employed to solve the above two problems, respectively. The RW-based pseudolabeling algorithm has been proved to be effective for high-confidence samples extraction \cite{Sun2017}, whereas C-CCA uses all pair-wise correspondences within a cluster across the two domains and results in cluster segregation 
\cite{Rasiwasia2014}. Fig. \ref{CCCA} illustrates the difference between CCA and C-CCA. It is clear that CCA requires paired samples and can hardly be directly applied when multiple clusters of samples in the source domain correspond to several clusters of samples in the target domain. 

\begin{figure*}[!t]
\setlength{\abovecaptionskip}{-5pt}
\setlength{\belowcaptionskip}{-20pt}
\centering
\includegraphics[width=0.95\textwidth]{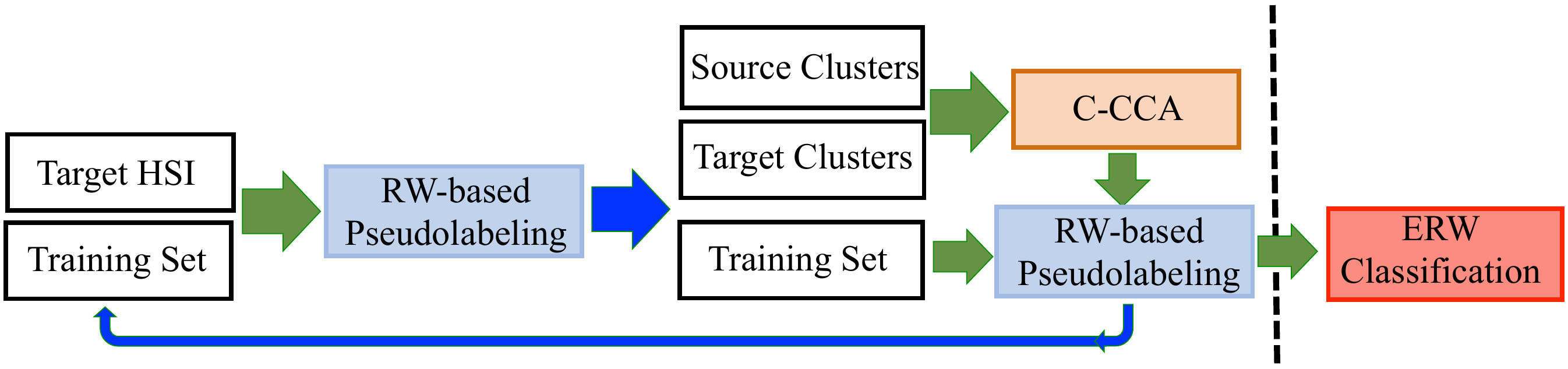}
\caption{Illustration of the proposed \textbf{CDCL} technique. Note that source clusters refer to the labeled source samples.}
\label{Flowchart}
\end{figure*}

In the proposed approach, the two algorithms work in a collaborative manner, i.e. RW-based pseudolabeling is employed to extract target samples with high confidence, whereas C-CCA is employed for cross-domain learning. Then the projected samples are used for RW-based pseudolabeling. Therefore, the proposed method is denoted as cross domain collaborative learning (\textbf{CDCL}). 
As is shown in Fig. \ref{Flowchart},  the proposed method is based on an iterative process, consisting of three main components, i.e. RW-based pseudolabeling, cross domain learning via C-CCA and classification using the extended RW (ERW) algorithm. Firstly, given the initially labeled target samples as training set ($\mathbf{TS}$), the RW-based pseudolabeling is employed to update the $\mathbf{TS}$ and extract target clusters ($\mathbf{TCs}$) by fusing the segmentation results obtained by RW and ERW classifiers. Secondly, cross domain learning via C-CCA is applied using labeled source samples and $\mathbf{TCs}$. The unlabeled target samples are then classified  with the estimated probability maps using the model trained in the projected correlation subspace. Then, both $\mathbf{TS}$ and estimated probability maps are used for updating $\mathbf{TS}$ again via RW-based pseudolabeling. Finally, when the iterative process converges, the classification map is obtained by the ERW classifier using the final $\mathbf{TS}$ and the estimated probability maps. 
Comprehensive experiments on four publicly available benchmark HSIs have been conducted to demonstrate the effectiveness of the proposed algorithm.

The rest of the paper is organized as follow. The C-CCA and RW algorithms are reviewed in Section II. The proposed methodology of \textbf{CDCL} is presented in section III. 
Section IV describes the experimental datasets and setup. Results and discussions are presented in Section V. Section VI summarizes the contributions of our research.

\section{Background Algorithms}
This section briefly describes background algorithms, i.e. C-CCA and RW algorithms.
\subsection{C-CCA}
Let us consider two sets of labeled samples  $\mathbf{X}^{s}_l$ and $\mathbf{X}^{t}_l$ extracted from the source and the target domains, respectively.  Each set is divided into $C$  corresponding clusters, which are denoted as $\{\mathbf{X}_{l,1}^s,...,\mathbf{X}_{l,C}^s\}$ and $\{\mathbf{X}_{l,1}^t,...,\mathbf{X}_{l,C}^t\}$, respectively. The $c$-th cluster of $\mathbf{X}^{s}_l$ is represented as $\mathbf{X}_{l,c}^s=\{\mathbf{x}_{1}^{s,c},...,\mathbf{x}_{|X_{l,c}^s|}^{s,c}\}$ and the $c$-th cluster of $\mathbf{X}^{l,t}$ is denoted as $\mathbf{X}_{l,c}^t=\{\mathbf{x}_{1}^{t,c},...,\mathbf{x}_{|X_{l,c}^t|}^{t,c}\}$, where $|X_{l,c}^s|$ and $|X_{l,c}^t|$ represent the number of samples in $\mathbf{X}_{l,c}^s$ and $\mathbf{X}_{l,c}^t$, respectively. The aim of C-CCA is to find a projection $\mathbf{u}$ for  $\mathbf{X}^{s}_l$ and $\mathbf{v}$
 for $\mathbf{X}^{t}_l$, so that correlation between  projections of $\mathbf{X}^{s}_l$ and $\mathbf{X}^{t}_l$ are maximized and clusters are well separated. 
  
In C-CCA, a one-to-one correspondence between all pairs of samples in a given cluster across the two sets is established and thereafter standard CCA is used to learn the projections. The C-CCA problem is written as
\begin{equation}
\rho = \max_{\mathbf{u},\mathbf{v}} \bigg\{ \frac{\mathbf{u}^{\mathrm{T}}\Sigma_{st}\mathbf{v}}{\sqrt{\mathbf{u}^{\mathrm{T}}\Sigma_{ss}\mathbf{u}}\sqrt{\mathbf{v}^{\mathrm{T}}\Sigma_{tt}\mathbf{v}}} \bigg\}
\end{equation}
where the covariance matrices $\Sigma_{st}$, $\Sigma_{ss}$ and $\Sigma_{tt}$ are defined as:
\begin{eqnarray}
\Sigma_{st}=\frac{1}{M}\sum_{c=1}^{C}\sum_{i=1}^{|X_{l,c}^s|}\sum_{j=1}^{|X_{l,c}^t|}\mathbf{x}_{i}^{s,c}(\mathbf{x}_{j}^{t,c})^{\mathrm{T}}\\
\Sigma_{ss}=\frac{1}{M}\sum_{c=1}^{C}\sum_{i=1}^{|X_{l,c}^s|}|X_{l,c}^t|\mathbf{x}_{i}^{s,c}(\mathbf{x}_{i}^{s,c})^{\mathrm{T}}\\
\Sigma_{tt}=\frac{1}{M}\sum_{c=1}^{C}\sum_{j=1}^{|X_{l,c}^t|}|X_{l,c}^s|\mathbf{x}_{j}^{t,c}(\mathbf{x}_{j}^{t,c})^{\mathrm{T}}
\end{eqnarray}
where $M=\sum_{c=1}^{C}|X_{l,c}^s||X_{l,c}^t|$ is the total number of cross-set correspondences. The problem can be solved as an eigenvalue problem as in CCA.

\subsection{RW}
The RW algorithm has been initially designed for general image segmentation based on a small set of labeled pixels \cite{Grady2006}. The algorithm assigns each unlabeled pixel to the label that a random walker starting from that pixel would be most likely to reach first.
To be specific, it considers an image as a graph \textbf{G}=$(V,E)$ with \textbf{vertices} $v\in V$ and \textbf{edges} $e\in E$. Then vertices and edges represent the pixels in the image and the links connecting the adjacent pixels, respectively. The structure of image intensities can be defined by the edge weights. The edge weight between the $i$-th and $j$-th pixels is defined as $w_{ij}=\mathrm{exp}(-\beta(g_i-g_j)^2)$, where $g_i$ indicates the image intensity at pixel $v_i$ and $\beta$ is a free parameter that controls the smoothness of graph edges.
The corresponding Laplacian matrix of the graph is denoted as $\mathbf{L}$.

The vertices $V$ of the image can be divided into  the set of labeled pixels $V_l$ and unlabeled set $V_u$, where $v_i\in V_l$ has assigned a label $c$ from the set $\mathcal{C}=\{1,2,...,C\}$.
Given the intensity representation of the image \textbf{G} and $V_l$, the RW algorithm determines the probability $\mathbf{p}_{ic}$ that a random walker starting at unlabeled pixel $i$ will first reach a labeled pixel belonging to $V_l$ with label $c$. 
The set of probabilities is addressed analytically and quickly with closed solutions by minimizing the energy function $E_{spatial}^{c}(\mathbf{p}_{c})=\mathbf{p}_{c}^{\mathrm{T}}\mathbf{L}\mathbf{p}_{c}$.
By assigning each pixel to the label with the largest probability, a high-quality image segmentation is obtained \cite{Grady2006}.

When RW is directly applied to HSI classification, the spectral information can hardly be integrated in the energy function $E_{spatial}^{c}$. 
To address the problem, an ERW-based spectral-spatial algorithm is proposed in \cite{Kang2015}, including the aspatial energy function defined as follows:
\begin{equation}
E_{aspatial}^{c}(\mathbf{p}_{c})=\sum_{q=1,q\neq c}^{C} \mathbf{p}_{q}^{\mathrm{T}}\Lambda_q\mathbf{p}_{q}+ (\mathbf{p}_{c}-1)^{\mathrm{T}}\Lambda_c(\mathbf{p}_{c}-1)
\end{equation}
where $\Lambda_c$ is a diagonal matrix, where the values are the initial probabilities for pixels.
The probabilities can be estimated by applying SVM classifier to the HSI image. The combined energy function for ERW algorithm is formulated as 
\begin{equation}
\label{Easpatial}
E^{c}(\mathbf{p}_c) = E_{spatial}^{c}(\mathbf{p}_{c}) + \gamma E_{aspatial}^{c}(\mathbf{p}_{c})
\end{equation}
where $\gamma$ is a free parameter controlling the dynamic range of the aspatial function. Similar to the solution of RW, the set of probabilities in ERW can be estimated by solving linear equations \cite{Grady2005}. Given the optimized probabilities, each unlabeled pixel is assigned with the label of the largest probability. 

\section{Proposed Method}

\subsection{Problem Definition}
Assume that we have $n_{s}$ labeled training samples $\{(\mathbf{x}_{i}^{s},y_i^s)|_{i=1}^{n_{s}}\}$ in the source domain, where $\mathbf{x}_{s}^{i} \in \mathbb{R}^{d_s}$ and $y_i^s\in\{1,2,...,C\}$. The samples in the target domain are divided into the labeled and unlabeled sets, which are denoted as $\{(\mathbf{x}_{i}^{tl},y_i^{tl})|_{i=1}^{n_{tl}}\}$ and $\{\mathbf{x}_{i}^{tu}|_{i=1}^{n_{tu}}\}$, respectively, and $y_i^{tl}\in\{1,2,...,C\}$.
In this paper, we only consider semi-supervised heterogeneous problem, thus we assume that $d_s\neq d_t$.  
For better illustration, the sets of labeled training source samples, the labeled and unlabeled target samples are denoted as $\mathbf{X}_{l}^{s}$, $\mathbf{X}_{l}^{t}$ and $\mathbf{X}_{u}^{t}$, respectively. 

As shown in Fig. \ref{Flowchart}, the proposed algorithm is based on an iterative process, including three main components, i.e. RW-based pseudolabeling, cross domain learning via C-CCA and ERW-based classification. It is notable that both RW-based pseudolabeling and  ERW-based classification require training set  and probability maps (which measure the probabilities that each sample of the target HSI belongs to different classes). 
The pseudolabeling procedure is introduced to extract several labeled samples with high-confidence as target clusters for C-CCA and more reliably labeled samples for updating the training set. To be specific, the strategy of RW-based label verification in \cite{Sun2017} is applied  to obtain reliable results of pseudolabeling. 
For simplicity, the estimated probability maps, target clusters and training set are denoted as $\widehat{\mathbf{P}}$, $\mathbf{TCs}$ and $\mathbf{TS}$, respectively.

In the following, the RW-based pseudolabeling will be first described. Then, the details of the proposed method will be introduced.

 \begin{figure}[!t]
\setlength{\abovecaptionskip}{-5pt}
\setlength{\belowcaptionskip}{-20pt}
\centering
\includegraphics[width=0.48\textwidth]{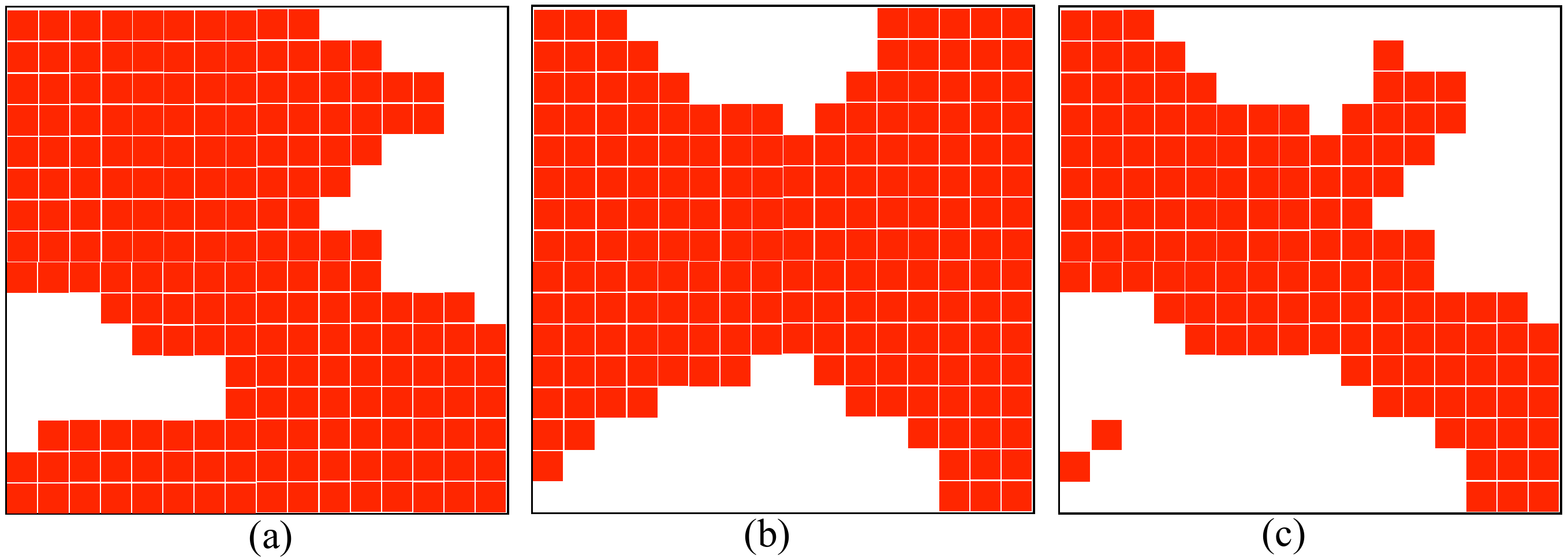}
\caption{Illustration of the RW-based label verification. (a) $\mathbf{S}_{rw}$: segmentation result of RW; (b) $\mathbf{S}_{erw}$: segmentation result of ERW; (c) Fusion of $\mathbf{S}_{rw}$ and $\mathbf{S}_{erw}$: samples with high confidences via label verification.}
\label{RW_PL}
\end{figure}

\subsection{RW-based Pseudolabeling}
Given the training set $\mathbf{TS}$ and the estimated probability map $\widehat{\mathbf{P}}$, the RW-based target samples pseudolabeling consists of the following five steps:

1) \emph{Graph construction}: In order to make full use of the spatial information, the first principal component (PC) of the hyperspectral image is used to construct a weighted graph \textbf{G}=$(V,E)$. Here, the \textbf{vertices} ($v\in V$) refer to the sample values in the first PC, and the \textbf{edges} ($e\in E$) refer to the links connecting the adjacent samples (eight neighbors are considered for each sample). A weight $w_{ij}=\mathrm{exp}(-\beta(v_i-v_j)^2)$ is defined for each edge $e_{ij}$ to model the difference between adjacent samples in the weighted graph, where $\beta$ is a free parameter.

2) \emph{RW segmentation}:  When the graph representation and $\mathbf{TS}$ are available, RW probabilities can be directly obtained by minimizing the energy function $E_{spatial}^{c}$. Denoted as $\mathbf{S}_{rw}$, the segmentation result is obtained by choosing the label with the maximum of probabilities for each sample. 

3) \emph{ERW segmentation}: Given the graph representation, $\mathbf{TS}$ and the initial probability map $\widehat{\mathbf{P}}$, 
ERW probabilities can be optimized by minimizing the energy function $E_{aspatial}^{c}$. Note that there is a free parameter $\gamma$ controlling the dynamic range of the aspatial function. Once the optimized probability map $\widehat{\mathbf{P}}_{erw}$ is obtained, the segmentation result $\mathbf{S}_{erw}$ is easily computed by choosing the label corresponding to the maximum of probabilities for each sample.

4) \emph{Label verification} \cite{Sun2017}: After obtaining $\mathbf{S}_{rw}$ and $\mathbf{S}_{erw}$, label verification is employed to extract sample candidates for further $\mathbf{TCs}$ and $\mathbf{TS}$ updating. As illustrated in Fig. \ref{RW_PL}, $\mathbf{S}_{rw}$ and $\mathbf{S}_{erw}$ are compared to verify the confidences of the unlabeled samples in the target HSI. To be specific, samples segmented as the same label in $\mathbf{S}_{rw}$ and $\mathbf{S}_{erw}$ are considered as the sample candidates with high confidences.
The rationality of the strategy is as follows. Firstly, the RW and ERW can take complementary decisions. The RW algorithm is only based on the spatial correlation among adjacent samples, whereas the ERW algorithm combines the spectral information and the spatial correlations of adjacent samples. Secondly, the core idea of the strategy is similar to the voting-based decision fusion strategy, i.e. if different classifiers take the same decision for a sample, the decision of this sample is assumed to be more reliable.

5) $\mathbf{TCs}$ and $\mathbf{TS}$ \emph{updating}:
Although candidate samples are extracted by label verification strategy with high confidences, $\mathbf{TS}$ and $\mathbf{TCs}$ are expected to include more correctly labeled samples. 
In order to ensure the accuracy of $\mathbf{TS}$, $p$ unlabeled samples in the candidate samples are selected according to the modified breaking ties (MBT)-based query strategy \cite{Li2011}. To be specific, the MBT strategy finds  the $p$ samples maximizing the ERW probability $\widehat{\mathbf{P}}$. Then the $p$ samples with the predicted label are added into $\mathbf{TS}$. In addition, the sample having its largest probability larger than the mean probability of its predicted class is used for $\mathbf{TCs}$ extraction.

\begin{displaymath}
\label{Detail}
\setlength{\abovedisplayskip}{2pt}
\setlength{\belowdisplayskip}{2pt}
\setlength{\arrayrulewidth}{0.4mm}
\begin{tabular}{p{8.2cm}}
\hline
$\! \mathbf{Algorithm \ 1}$ Cross Domain Collaborative Learning \\
\hline
$\mathbf{Input}$: \\
\ \ \ Samples $\mathbf{X}_{l}^{s}$, $\mathbf{X}_{l}^{t}$ and $\mathbf{X}_{u}^{t}$. \\
\ \ \ Threshold for correlation coefficients: $\rho_T$=0.5.\\    
\ \ \ Initial training set: $\mathbf{TS} = \mathbf{X}_{l}^{t}$.\\
\ \ \ Free parameters for ERW: $\beta =710,\gamma=1e-5$.\\
\ \ \ Number of samples added into $\mathbf{TS}$: $p=10$.\\
1: \textbf{Repeat}: \\
2: \ \ Train a linear SVM classifier with probability esti-\\
\ \ \ \ \ mation using $\mathbf{TS}$.\\
3: \ \ Classify $\mathbf{X}_{u}^{t}$ and estimate probability map $\widehat{\mathbf{P}}_1$.\\
4: \ \ Update $\mathbf{TS}$ and $\mathbf{TCs}$ via RW-based pseudolabeling \\
\ \ \ \ \ using $\mathbf{TS}$ and $\widehat{\mathbf{P}}_1$.\\
5: \ \ C-CCA using $\mathbf{X}_{l}^{s}$ and $\mathbf{TCs}$.\\
6: \ \ Samples projected onto the subspace kept by $\rho_T$.\\
7: \ \ Train a linear SVM classifier with probability esti- \\
\ \ \ \ \ mation using projected $\mathbf{X}_{l}^{s}$ and $\mathbf{X}_{l}^{t}$.\\
8: \ \ Classify $\mathbf{X}_{u}^{t}$ and estimate probability map $\widehat{\mathbf{P}}_2$.\\
9: \ \ Update $\mathbf{TS}$ via RW-based pseudolabeling using $\mathbf{TS}$ \\
\ \ \ \ \ and $\widehat{\mathbf{P}}_2$.\\
10: \textbf{Until convergence}\\
11: ERW-based classification using $\mathbf{TS}$ and $\widehat{\mathbf{P}}_2$.\\
12: \textbf{Return} Classification map\\
\hline
\end{tabular}
\end{displaymath}

\subsection{Details of Proposed Technique}

As illustrated in Algorithm 1, the proposed algorithm can be denoted as cross domain collaborative learning (\textbf{CDCL}) via RW-based pseudolabeling and C-CCA, with $\mathbf{TS}$ and $\mathbf{TCs}$ updated iteratively. The details of the proposed algorithm are as follows.

1) \emph{RW-based pseudolabeling}: As illustrated in Fig. \ref{Flowchart}, in one iterative process, pseudolabeling is applied twice, i.e. before and after C-CCA. Firstly, probability estimation for pseudolabeling is achieved by training a linear SVM classifier on $\mathbf{TS}$.
Then, the obtained probability maps $\widehat{\mathbf{P}}_1$  and $\mathbf{TS}$ are employed to extract $\mathbf{TCs}$ and update $\mathbf{TS}$. Note that the initial $\mathbf{TS}$ only contains $\mathbf{X}_{l}^{t}$.
Secondly, after C-CCA using $\mathbf{TCs}$ and $\mathbf{X}_{l}^{s}$, the probability maps are estimated by the linear SVM trained using the projected $\mathbf{X}_{l}^{s}$ and $\mathbf{X}_{l}^{t}$. Given $\mathbf{TS}$ and the newly estimated probability map $\widehat{\mathbf{P}}_2$, pseudolabeling is applied again for updating $\mathbf{TS}$. In summary, $\mathbf{TS}$ is updated twice and $\mathbf{TCs}$ is computed only once in a single iterative process. 

2) \emph{Cross domain learning via C-CCA}: Given $\mathbf{X}_{l}^{s}$ and $\mathbf{TCs}$, more than one pair of projection vectors $\{\mathbf{u}_{i}^{s}\}_{i=1}^d$ and $\{\mathbf{v}_{i}^{t}\}_{i=1}^d$ with corresponding correlation coefficient $\rho_i$ are derived via C-CCA. Note that $d$ is the dimension number of the obtained subspace, which is smaller than both $d_s$ and $d_t$. Higher values of the correlation coefficients indicate better correlations between samples projected from different domains, resulting in better domain transfer abilities. In order to generalize correlation subspace with good transfer abilities, we fix the threshold for $\rho$ as 0.5 and the corresponding vectors are kept. After projecting all samples in both domains onto the correlation subspace, $\mathbf{X}_{u}^{t}$ are classified with estimated probability maps $\widehat{\mathbf{P}}_2$ by using a linear SVM trained on the projected $\mathbf{X}_{l}^{s}$ and $\mathbf{X}_{l}^{t}$. Although non-linear classifiers like SVM with RBF kernel generally perform better than linear classifiers in classification task, the optimal parameters of such classifier tuned by source samples usually perform worse than expected for target samples under the context of DA. On the contrary, linear kernel can capture original relationships between samples from different domains.

3) \emph{ERW-based classification}: When the iterative process of RW-based pseudolabeling and C-CCA converges, the classification map is obtained by ERW using the estimated probability maps  $\widehat{\mathbf{P}}_2$ and the final $\mathbf{TS}$.

\subsection{Performance Analysis and Convergence}
The classification ability and convergence of the proposed \textbf{CDCL} method are analyzed as follows: 

1) Given $\mathbf{X}_{l}^{s}$ and $\mathbf{X}_{l}^{t}$, the classification ability of the proposed method relies on two factors, i.e. transfer abilities of C-CCA and ERW-based classification. 
It is clear that the transfer abilities of C-CCA relies on the number of samples in $\mathbf{TCs}$ and the corresponding accuracies, whereas ERW-based classification requires a good estimation of $\mathbf{P}$ and $\mathbf{TS}$ to achieve higher accuracy.
In each iterative process, the $q$ samples with highest confidence are added into $\mathbf{TS}$ and several samples are extracted by label verification as $\mathbf{TCs}$. If $\mathbf{TS}$ and $\mathbf{TCs}$ are accurate, good cross domain learning would be achieved. Since the classifier is trained using labeled samples from both domains, $\widehat{\mathbf{P}}_2$ performs better than $\widehat{\mathbf{P}}_1$ when it is used for RW-based pseudolabeling. Therefore, more reliable samples are added into $\mathbf{TS}$, ensuring $\mathbf{Ts}$ and $\mathbf{TCs}$ are accurately updated in the next iteration. With $\mathbf{TS}$ and $\mathbf{TCs}$ updated iteratively, good classification result can be obtained using the proposed method.

2) As stated above, higher classification accuracy via the proposed method is easily obtained under the assumption to have reasonably accurate $\mathbf{TS}$ and $\mathbf{TCs}$. 
However, note that since the C-CCA is based on the pairwise correspondences within a cluster across domains, it is expected that the source clusters are aligned with the corresponding target clusters even if there are few mislabelled samples in $\mathbf{TCs}$.
With the iterative process going on,  both RW and ERW segmentation results will be close to the ground truth, resulting in more samples extracted as candidates via label verification. Then more samples would be extracted as $\mathbf{TCs}$. Since samples with their probability larger than the mean probabilities of their predicted class are considered as  $\mathbf{TCs}$, the number of samples in $\mathbf{TCs}$ is smaller than the number of all unlabeled samples. In fact, the number of samples in $\mathbf{TCs}$ can hardly be monotonically increasing with iterations due to the inconsistency between segmentation results obtained by RW and ERW algorithms. Therefore, if the increase  of sample amount in $\mathbf{TCs}$ is less than 5\% of the total unlabeled samples, we consider the convergence reached. 

\section{Experimental Data and Setup}

\subsection{DataSet Description}
\definecolor{class1_1}{RGB}{192,192,192}
\definecolor{class1_2}{RGB}{0,255,0}
\definecolor{class1_3}{RGB}{0,255,255}
\definecolor{class1_4}{RGB}{0,128,0}
\definecolor{class1_5}{RGB}{255,0,255}
\definecolor{class1_6}{RGB}{165,82,41}
\definecolor{class1_7}{RGB}{128,0,128}
\definecolor{class2_1}{RGB}{140,67,46}
\definecolor{class2_2}{RGB}{0,0,255}
\definecolor{class2_3}{RGB}{255,100,0}
\definecolor{class2_4}{RGB}{0,255,123}
\definecolor{class2_5}{RGB}{164,75,155}
\definecolor{class2_6}{RGB}{101,174,255}
\definecolor{class2_7}{RGB}{118,254,172}
\definecolor{class2_8}{RGB}{61,91,112}
\definecolor{class2_9}{RGB}{255,255,0}
\definecolor{class2_10}{RGB}{255,255,125}
\definecolor{class2_11}{RGB}{255, 0, 255}
\definecolor{class2_12}{RGB}{100,0,255}
\definecolor{class2_13}{RGB}{0,172,254}
\definecolor{class2_14}{RGB}{0,255,0}
\definecolor{class2_15}{RGB}{171,175,80}
\definecolor{class2_16}{RGB}{101,193,60}
\begin{table}[!t]
\setlength{\abovecaptionskip}{0pt}
\setlength{\belowcaptionskip}{0pt}
\caption{Number of Labeled Samples Available for Pavia Data Set (Top) and Salinas/Indian Data Set (Down).}
\label{Dataset_Class}
{\begin{tabular}{l@{}lccccc@{}}
\hline 
\multirow{2}{*}{No.} & \multirow{2}{*}{ Class }& \multirow{2}{*}{Color}  & \multicolumn{2}{c}{Pavia University }& \multicolumn{2}{c}{Pavia Center} \\
\cline{4-7}
{}&{}&{}&{TM}&{GT}&{TM}&{GT}\\
\hline\hline
1     & Asphalt& \cellcolor{class1_1}        & 548 &6631   &678    &7585   \\ 
2     & Meadows& \cellcolor{class1_2}    &540  & 18649  &797       &2905 \\ 
3    & Trees& \cellcolor{class1_3}    & 524 & 3064     &  785     &6508 \\ 
4     & Baresoil& \cellcolor{class1_4}    & 532  &5029   &  820     &6549  \\
5   & Bricks& \cellcolor{class1_5}    & 514  &3682       &  485    &2140 \\
6    &Bitumen & \cellcolor{class1_6}    & 375  &1330   &  808   &7287  \\
7    &Shadows & \cellcolor{class1_7}    & 231  &947     &  195       &2165  \\
\hline
{No.} & { Class }&  {Color}  & \multicolumn{2}{c}{Salinas}& \multicolumn{2}{c}{Indian}\\
\hline\hline
1     &Weeds\_1/Alfalfa & \cellcolor{class2_1}    & \multicolumn{2}{c}{2009}    &\multicolumn{2}{c}{46}  \\
2     &Weeds\_1/Corn\_n& \cellcolor{class2_2}    & \multicolumn{2}{c}{3726}     &\multicolumn{2}{c}{1428} \\
3     &Fallow/Corn\_m& \cellcolor{class2_3}    & \multicolumn{2}{c}{1976}     &\multicolumn{2}{c}{830}\\
4     &Fallow\_r/Corn& \cellcolor{class2_4}    & \multicolumn{2}{c}{1394}    &\multicolumn{2}{c}{237}\\
5     &Fallow\_s/Grass-pasture& \cellcolor{class2_5}    & \multicolumn{2}{c}{2678}     &\multicolumn{2}{c}{483}\\
6     &Stubble/Grass-trees& \cellcolor{class2_6}    & \multicolumn{2}{c}{3959}     &\multicolumn{2}{c}{730} \\
7     &Celery/Grass-pasture\_m& \cellcolor{class2_7}    &\multicolumn{2}{c}{3579}     &\multicolumn{2}{c}{28}\\
8     &Graphes\_u/Hay\_w& \cellcolor{class2_8}    &\multicolumn{2}{c}{11271}    &\multicolumn{2}{c}{478}\\
9     &Soil\_v/Oats& \cellcolor{class2_9}    & \multicolumn{2}{c}{547}    &\multicolumn{2}{c}{20}  \\
10     &Corn\_s/Soybean\_n& \cellcolor{class2_10}    &\multicolumn{2}{c}{3278}     &\multicolumn{2}{c}{972} \\
11     &Lettuce\_4wk/Soybean\_m& \cellcolor{class2_11}    &\multicolumn{2}{c}{1068}     &\multicolumn{2}{c}{2455}\\
12     &Lettuce\_5wk/Soybean\_c& \cellcolor{class2_12}    &\multicolumn{2}{c}{1927}     &\multicolumn{2}{c}{593}\\
13     &Lettuce\_6wk/Wheat& \cellcolor{class2_13}    &\multicolumn{2}{c}{916}     &\multicolumn{2}{c}{205}\\
14     &Lettuce\_7wk/Woods& \cellcolor{class2_14}    &\multicolumn{2}{c}{1070}     &\multicolumn{2}{c}{1265} \\
15     &Vinyard\_u/Buildings-Grass& \cellcolor{class2_15}    &\multicolumn{2}{c}{7268}     &\multicolumn{2}{c}{386}\\
16     &Vinyard\_v/Stone-Steel& \cellcolor{class2_16}    &\multicolumn{2}{c}{1807}    &\multicolumn{2}{c}{93}\\
\hline
\end{tabular}}
\end{table}

\begin{figure}[!t]
\setlength{\abovecaptionskip}{-5pt}
\setlength{\belowcaptionskip}{-20pt}
\centering
\includegraphics[width=0.48\textwidth]{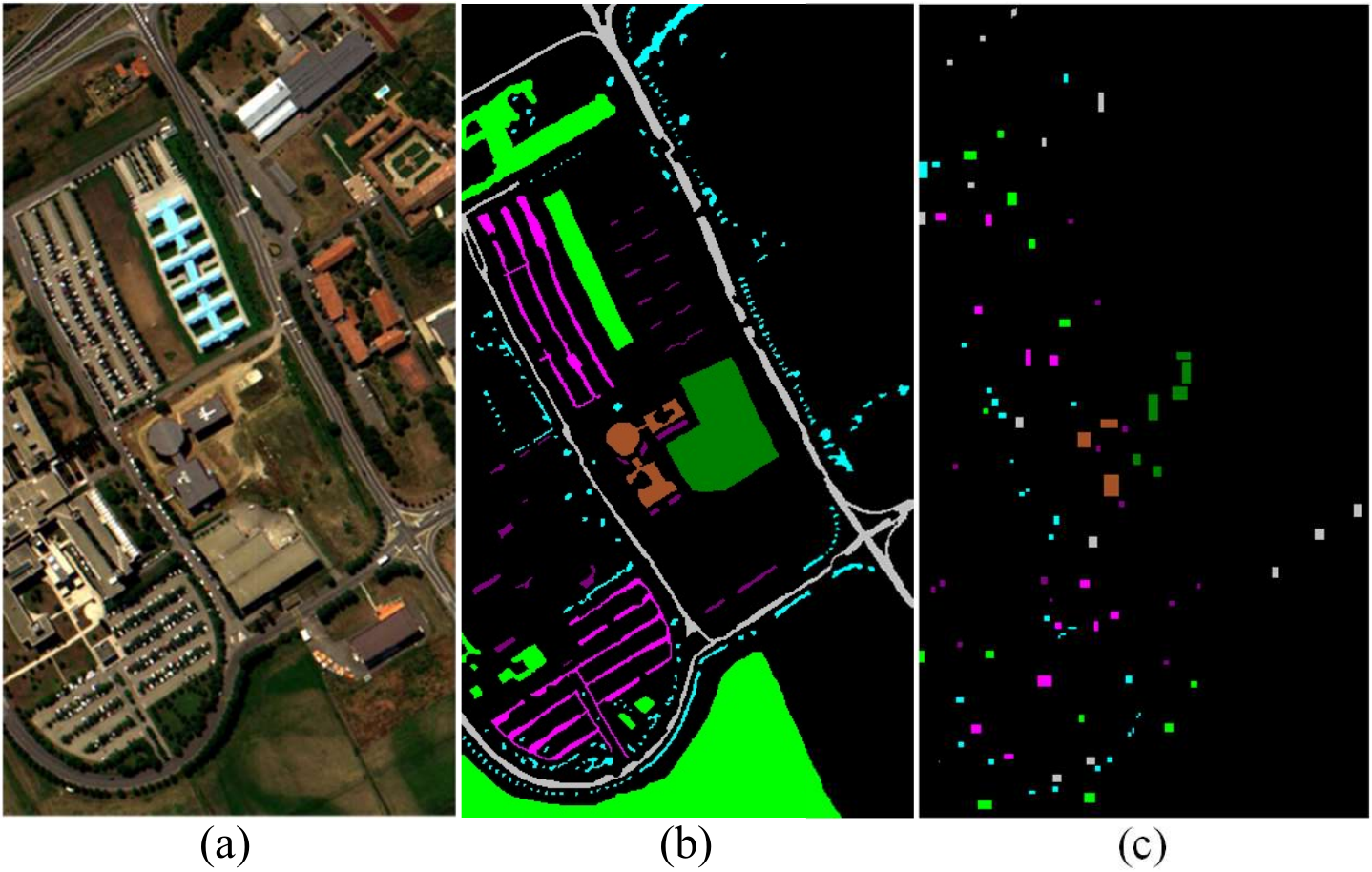}
\includegraphics[width=0.48\textwidth]{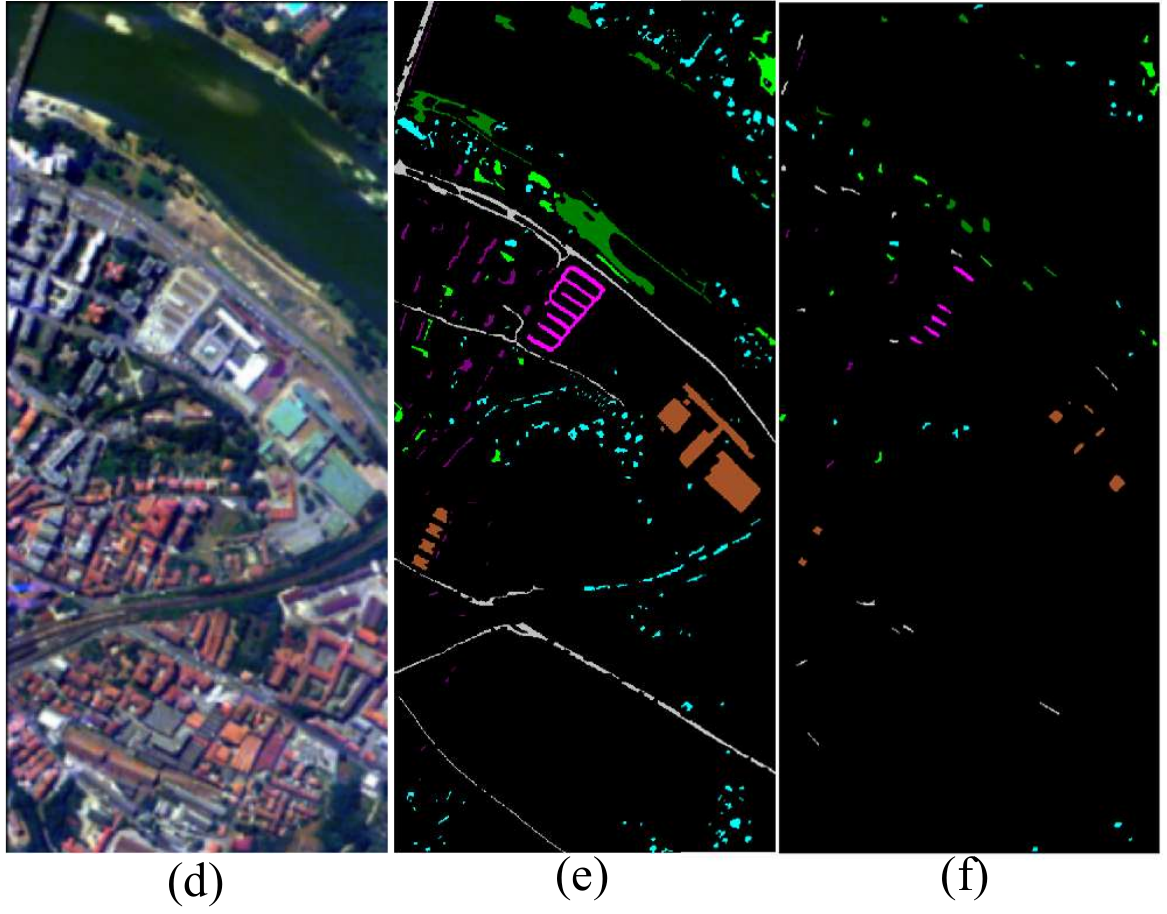}
\caption{ROSIS Pavia dataset used in our experiments. (a) Color composite image, (b) ground truth and (c) training map of the University scene; (d) color composite image, (e) ground truth and (f) training map of City Center scene.}
\label{Pavia}
\end{figure}

The first dataset consists of two hyperspectral images collected by the Reflective Optics Spectrographic Image System (ROSIS) sensor over the University of Pavia and Pavia City Center. The Pavia City Center image contains 102 spectral bands and has a size of 1096$\times$492 pixels. The Pavia University image contains instead 103 spectral reflectance bands and has a size of 610$\times$340 pixels. Only seven classes shared by both images are considered herein. In the experiments, the Pavia University image is considered as the source domain, while the Pavia City Center image as the target domain, or vice versa. These two cases are denoted as \emph{Univ/Center} and \emph{Center/Univ}, respectively. Note that there are manually selected training maps (TM) which are publicly available and widely used in related publications \cite{Li2011,Sun2017,Fauvel2013,Tarabalka2010}. The color composite image,  ground truth (GT) and TM of Pavia dataset are illustrated in Fig. \ref{Pavia}, whereas the corresponding number of labeled samples is detailed in Table \ref{Dataset_Class}.
  
\begin{figure}[!t]
\setlength{\abovecaptionskip}{-5pt}
\setlength{\belowcaptionskip}{-20pt}
\centering
\includegraphics[width=0.48\textwidth]{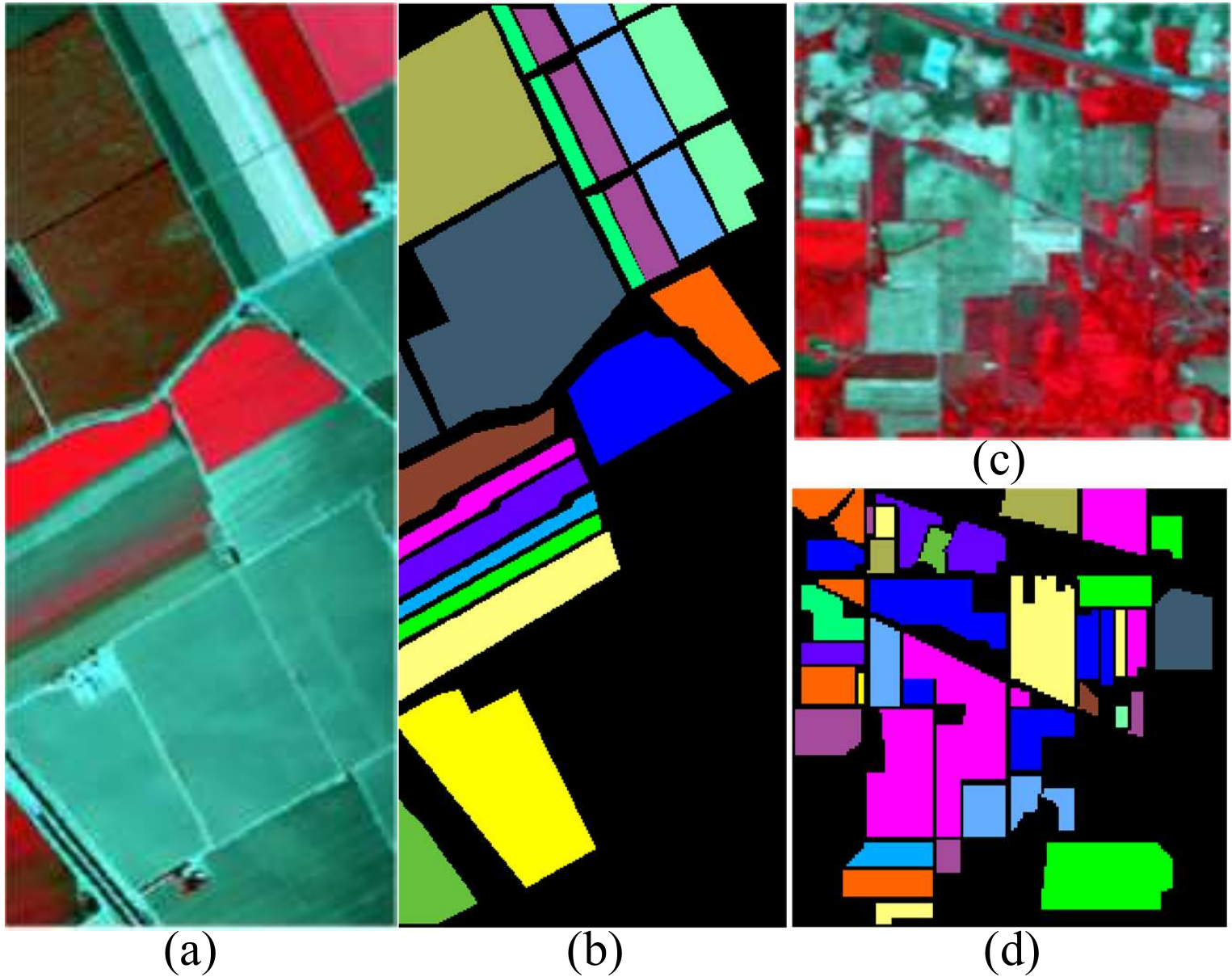}
\caption{ AVIRIS Salinas and Indian Pines datasets used in our experiments. (a) Color composite image and (b) ground truth of the Salinas data; (c) color composite image and (d) ground truth of Indian Pines data.}
\label{Salinas/Indian}
\end{figure}

The second dataset consists of two hyperspectral images captured with Airborne Visible Infrared Imaging Spectrometer (AVIRIS) over Salinas Valley, California and Northwest Indiana. After discarding 20 water absorption bands, Salinas image contains 224 bands of 512 $\times$ 217 pixels. Fig. \ref{Salinas/Indian}(a-b) show the color composite image and the GT of the Salinas data set, in which 16 different classes represent mostly different types of crops.
 After removing 20 spectral bands due to noise and water absorption, Indian Pines image contains 200 bands of 145 $\times$145 pixels, and its spatial resolution is 20 m per pixel. The color composite image and the GT containing 16 different classes are presented in Fig. \ref{Salinas/Indian}(c-d). 
The classes of both images are listed in Table \ref{Dataset_Class} with the corresponding number of samples. Since we mainly focus on the HDA problem, a low-dimensional image is considered as the source domain by clustering the spectral space of the original data for each image. Specifically, the original bands of the HSI are clustered into 50 groups using the K-means algorithm, and the mean value of each cluster is considered as a new spectral band, providing a total of 50 new bands. The corresponding cases are denoted as \emph{Salinas} and \emph{Indian} cases, respectively.

\subsection{Experimental Setup}
In order to make a general comparison, the default parameters of the ERW classifier given in \cite{Kang2015,Sun2017} are adopted for the proposed algorithm. Specifically, the parameters of the RW and ERW in the proposed method are set to be $\beta=710$ and $\gamma=1e-5$. In addition, the threshold of correlation coefficients $\rho_T$ and the query size $q$ are set to 0.5 and 10, respectively, in all experiments. 
The free parameter $C$ of linear SVM in our method is tuned in the range ($2^{-3}-2^{10}$) with 5-fold cross-validation.

Several approaches of semi-supervised HDA proposed for visual and remote sensing applications are  employed as baseline methods:

$\bullet$ \textbf{CCA} \cite{Hardoon2004}: \emph{CCA} aligns both domains by using the same number of labeled samples from source and target domains. To be specific, a random selection of samples from source or target domain is applied to ensure pairwise correspondences between domains. 

$\bullet$ \textbf{C-CCA} \cite{Rasiwasia2014}: \emph{C-CCA} is directly employed by using the labeled samples in both domains.

$\bullet$ \textbf{DAMA} \cite{Wang2011}:  \emph{DAMA} adopts a linear projection to match the differences between the source and target subspaces.

$\bullet$ \textbf{SSMA} \cite{Tuia2014}: \emph{SSMA} carries out adaptation through manifold alignment while preserving label (dis)similarities and the geometric structures of the single manifold in both domains.

$\bullet$ \textbf{KEMA} \cite{Tuia2016}: \emph{KEMA} is a kernerlized version of SSMA. 

$\bullet$ \textbf{SHFA} \cite{Li2014}: \emph{SHFA} simultaneously learns the target classifier and infers the target labels in an augmented common feature space.

$\bullet$ \textbf{CDLS} \cite{HubertTsai2016}: \emph{CDLS} jointly explores a domain-invariant feature subspace and identifies cross-domain landmarks.
\begin{table}[!t]
\caption{Number of Training and Test Samples Used for the Pavia and Salinas/Indian Data Sets. }
\label{Setup}
{\begin{tabular}{l|c|c|c|c}
\hline
Data Set&\emph{Univ/Center}&\emph{Center/Univ}&\emph{Salinas/Salinas}&\emph{Indian/Indian}\\
\hline 
TR\_S & 50&10, 20, 50&50&5, 10, 15\\
TR\_T & 2&2, 3, 5&2&2, 3, 5\\
TE\_T & 2\%&2\%&2\%&10\%\\
\hline 
\end{tabular}}
\end{table}

\begin{table*}[!t]
\setlength{\abovecaptionskip}{0pt}
\setlength{\belowcaptionskip}{0pt}
\caption{Classification results for the Pavia Center Dataset. The Best Results For Each Row are Reported in Italic Bold. The Proposed \textbf{CDCL} Approach Significantly Outperforms All the Baseline Methods.}
\label{result_city}
{\begin{tabularx}{1\textwidth}{c|X|X|X|X|X|X|X|X|X|X|X}
\toprule[2pt]
\multirow{2}{*}{Class} &\multicolumn{11}{c}{Methods}\\
\cline{2-12}
{}  & \emph{CCA} & \emph{C-CCA}  & \emph{DAMA} & \emph{SSMA} & \emph{KEMA} & \emph{SHFA} & \emph{CDLS} & \emph{NA} & \emph{LapSVM} & \emph{ERW} & \textbf{CDCL}\\
\midrule[1pt]
\emph{Asphalt}& $43.55/3.4$& $84.80/2.4$& $91.18/3.2$& $94.70/1.8$& $96.02/0.5$& $79.21/2.7$& $39.41/5.6$& $94.28/0.9$& $76.55/4.3$& $\textbf{98.32}/1.1$& $95.13/1.6$\\ 
\emph{Meadows}& $46.02/5.8$& $69.92/3.4$& $80.17/3.4$& $75.85/4.8$& $74.66/3.3$& $83.64/2.8$& $39.07/5.1$& $77.71/4.2$& $\textbf{82.29}/2.7$& $58.39/7.4$& $59.75/4.7$\\ 
\emph{Trees}& $46.56/4.7$& $76.18/3.2$& $70.11/4.8$& $72.82/5.3$& $72.06/3.9$& $72.82/3.2$& $\textbf{92.63}/1.8$& $73.51/4.1$& $75.61/3.3$& $54.73/8.7$& $77.21/4.6$\\ 
\emph{Baresoil}& $38.21/3.0$& $53.51/3.8$& $40.34/5.2$& $47.29/5.3$& $58.51/3.9$& $72.86/3.2$& $58.36/4.0$& $67.75/3.4$& $66.56/2.8$& $70.95/7.1$& $\textbf{76.95}/4.7$\\ 
\emph{Bricks}& $36.16/4.2$& $60.81/3.5$& $73.02/4.1$& $72.44/4.2$& $70.47/2.5$& $66.74/3.9$& $53.95/5.9$& $81.63/2.5$& $59.30/4.4$& $81.63/5.7$& $\textbf{96.86}/2.3$\\ 
\emph{Bitumen}& $40.65/3.3$& $68.97/2.4$& $57.81/5.4$& $64.18/3.9$& $63.22/3.6$& $75.51/1.8$& $79.69/3.8$& $69.49/3.4$& $75.14/1.9$& $69.73/5.0$& $\textbf{82.50}/2.8$\\ 
\emph{Shadows}& $74.32/3.7$& $96.93/0.8$& $\textbf{100}/0.0$& $\textbf{100}/0.0$& $99.89/0.1$& $57.95/4.4$& $64.20/2.6$& $\textbf{100}/0.0$& $90.34/2.4$& $92.39/1.4$& $99.43/0.4$\\ 
\midrule[1pt]
OA&$44.19/1.8$& $72.17/1.4$& $69.46/1.5$& $72.93/1.2$& $74.73/1.1$& $74.37/1.1$& $63.53/1.2$& $78.58/1.2$& $74.52/1.2$& $74.52/2.2$& $\textbf{83.24}/1.5$\\ 
AA&$51.46/1.4$& $72.11/1.2$& $74.22/1.1$& $76.99/0.8$& $76.33/0.7$& $76.44/0.8$& $68.11/1.0$& $79.81/0.7$& $74.64/1.0$& $77.41/2.0$& $\textbf{82.29}/1.2$\\ 
Kappa&$34.22/0.2$& $66.91/0.2$& $63.96/0.2$& $67.91/0.1$& $69.87/0.1$& $69.47/0.1$& $56.23/0.1$& $74.55/0.1$& $69.71/0.1$& $69.68/0.3$& $\textbf{80.00}/0.2$\\ 
\bottomrule[2pt]
\end{tabularx}}
\end{table*}

Moreover, several methods applied only to the target domain are also employed as baselines:

$\bullet$ \textbf{No Adaptation} (NA): \emph{NA} is a basic baseline that learns linear SVM \cite{Chang2011} using the initially labeled target samples. 

$\bullet$ \textbf{LapSVM} \cite{Melacci2011}: \emph{LapSVM} is a typical baseline for semi-supervised classification and the one-vs-one strategy for linear SVM is applied for fair comparison. 

$\bullet$ \textbf{ERW} \cite{Kang2015}: \emph{ERW} carries out classification using the initial probabilities learned by linear SVM and the initially labeled target samples.\\
The threshold of correlation coefficient for \emph{CCA} and \emph{C-CCA} is set as 0.5. The parameter $\mu$ is set as 0.9 for \emph{DAMA}, \emph{SSMA} and \emph{KEMA}, whereas the optimal dimensionality of final projection for the three methods are cross-validated by exploiting labeled source and target samples. Once samples are projected onto new subspace, the final classification results of \emph{CCA}, \emph{C-CCA}, \emph{DAMA}, \emph{SSMA} and \emph{KEMA} are obtained by training the linear SVM using labeled samples of both domains with parameter $C$ tuned in the range ($2^{-3}-2^{10}$). The parameters of SHFA is tuned as in \cite{Li2014}.
The dimensionality of PCA in \emph{CDLS} is set as 30, whereas other parameters of \emph{CDLS} are tuned as in \cite{HubertTsai2016}. The parameters of \emph{LapSVM} are set to be $\gamma_I=1e-5$ and $\gamma_A=0.5$, whereas parameters of 
\emph{ERW} are set to be $\beta = 710$ and $\gamma = 1e-5$ for fair comparison.

\begin{figure}
\setlength{\abovecaptionskip}{-5pt}
\setlength{\belowcaptionskip}{-20pt}
\centering
\includegraphics[width=0.48\textwidth]{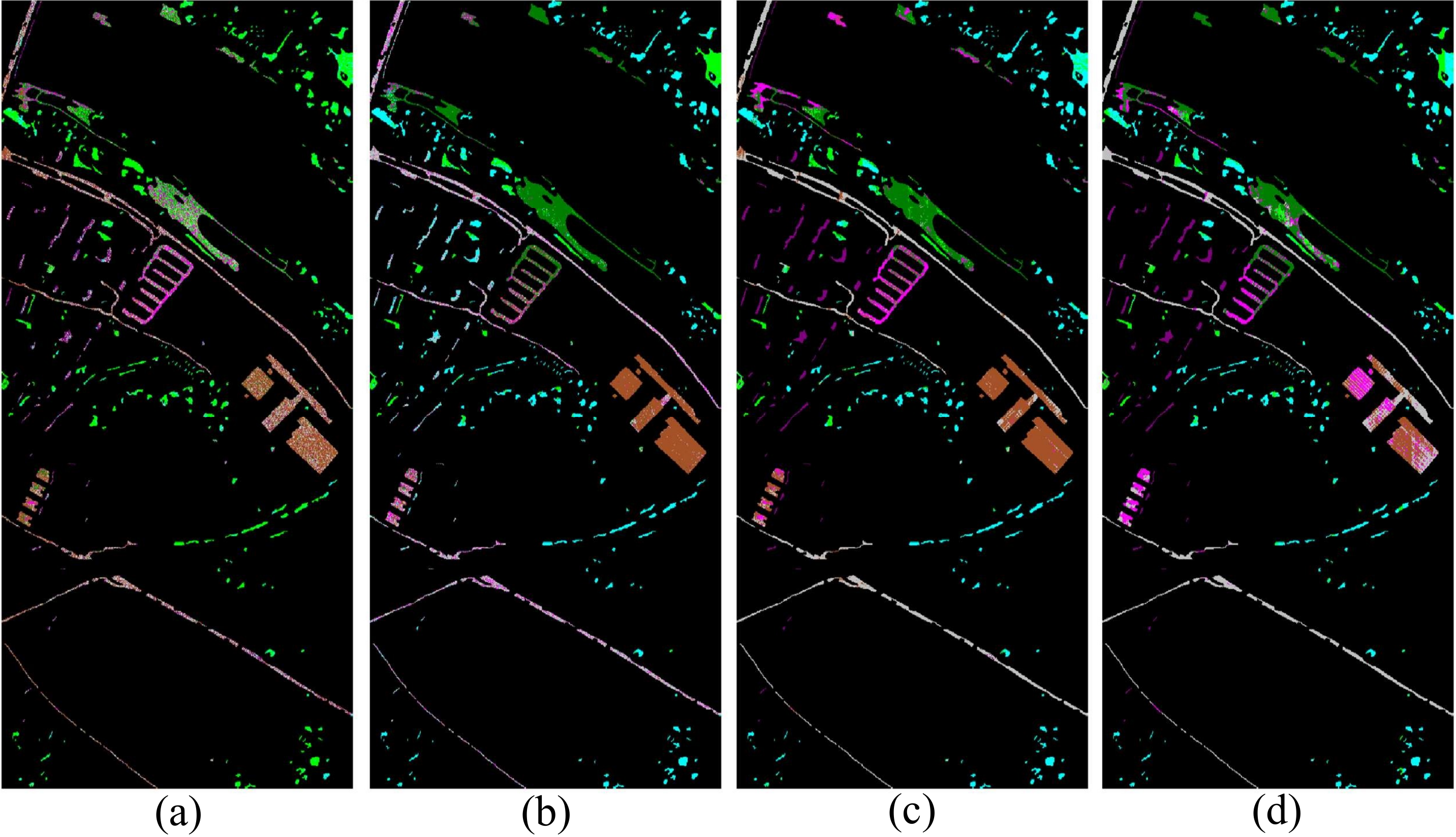}
\includegraphics[width=0.48\textwidth]{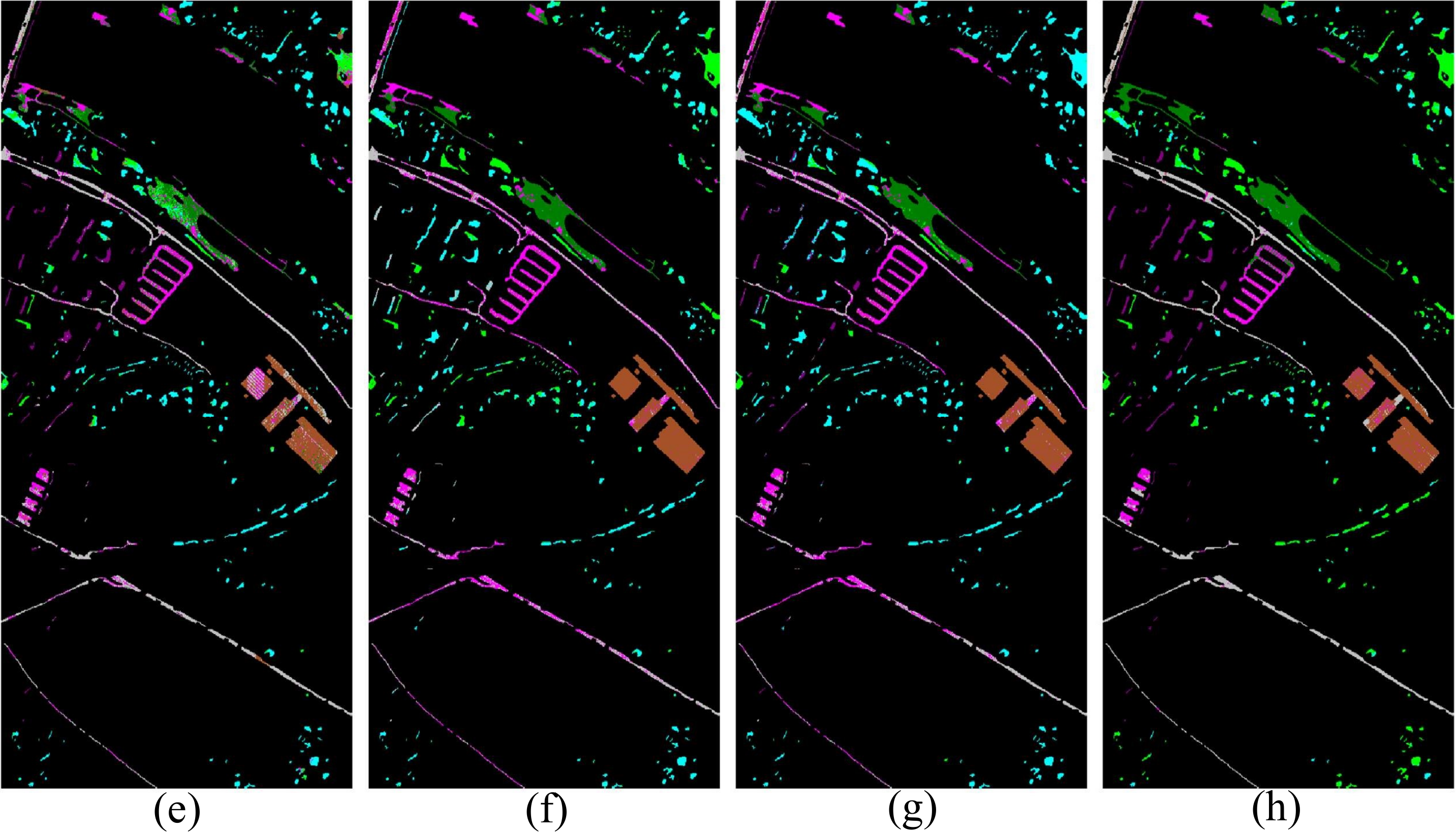}
\includegraphics[width=0.48\textwidth]{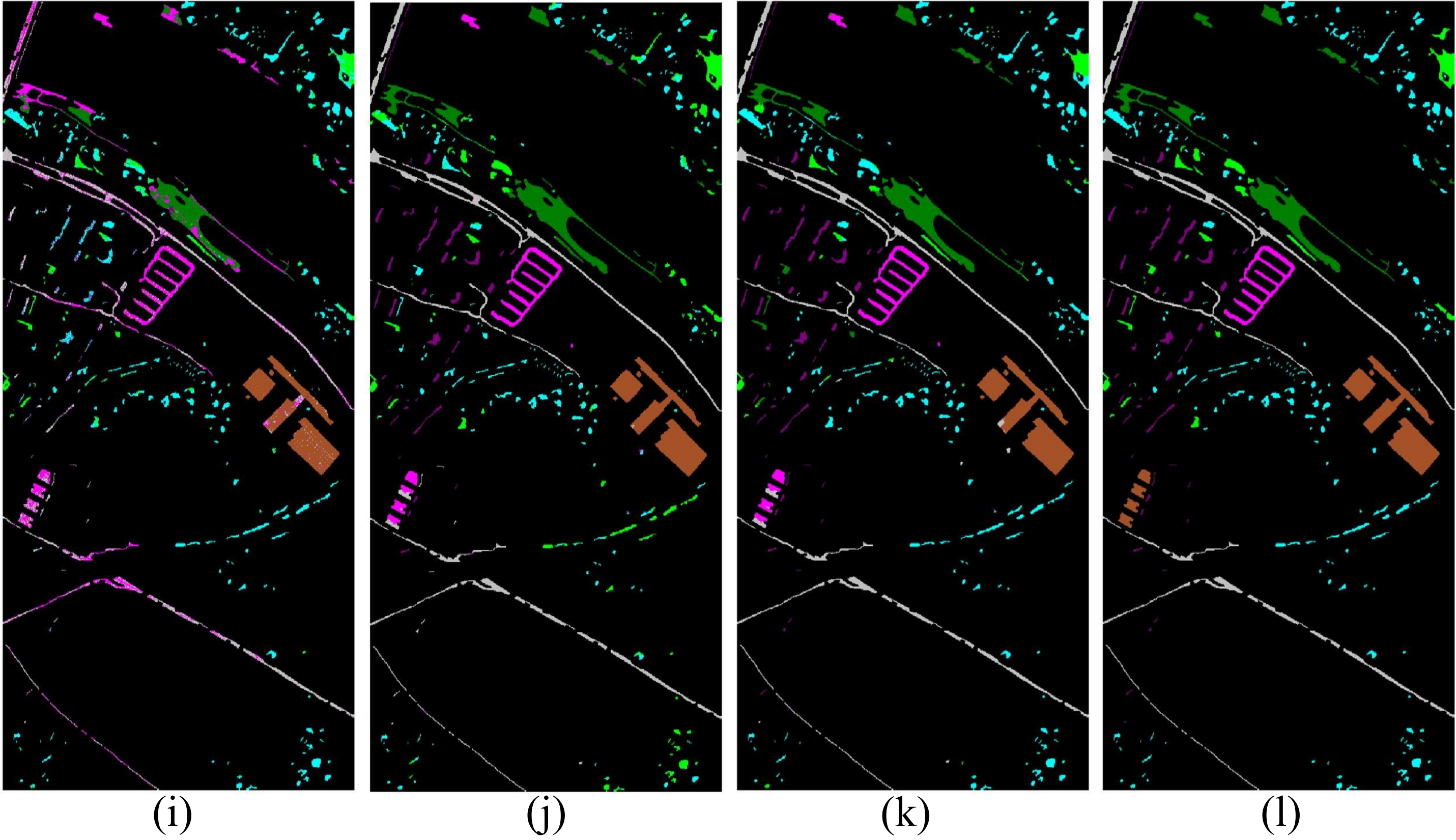}
\caption{Classification map of Pavia Center by (a)  \emph{CCA} (OA=44.63\%), (b) \emph{C-CCA} (OA=75.24\%),  (c) \emph{DAMA} (OA=82.28\%), (d) \emph{SSMA} (OA=71.51\%),  (e) \emph{KEMA} (OA=72.26\%), (f) \emph{SHFA} (OA=69.31\%),  (g) \emph{CDLS} (OA=70.93\%), (h) \emph{NA} (OA=79.98\%),  (i) \emph{LapSVM} (OA=71.74\%), (j) \emph{ERW} (OA=85.89\%),  (k) \emph{CDCL} (OA=91.03\%) methods and (l) denotes the corresponding ground truth.}
\label{CenterMap}
\end{figure}

In a practical application, the number of labeled samples in the target HSI is typically not enough to learn a reliable classifier, whereas the amount of labeled samples in the source HSI is relatively larger. To model this scenario, we randomly select a limited amount of samples from the target HSI as labeled. Table \ref{Setup} lists the settings of training and test samples used in our experiments, which consist of three parts: 1) training samp les (labeled) from the source HSI (TR\_S); 2) training samples (labeled) from the target HSI (TR\_T); and 3) test samples (unlabeled) from the target HSI (TE\_T). The integers (i.e., 2 3 5) in Table \ref{Setup} represent the number of samples per class, whereas the percentages refer to the ratio of training or testing samples. For example, the setting of \emph{Univ/Center} case means that 50 labeled source samples and 2 labeled target samples per class are selected as training samples, and 2\% of all unlabeled target samples are used for testing. Note that testing samples of four cases are selected from the corresponding target ground truth. The training samples for Pavia dataset are selected from publicly available training maps (see Fig. \ref{Pavia}), whereas the training samples for Salinas and Indian datasets are selected from the ground truth. To exploit the effectiveness of various training samples in both domains, various settings of TR\_S and TR\_T are applied for \emph{Center/Univ} and \emph{Indian} cases. For each setting in Table \ref{Setup}, 50 trials of the classification have been performed to ensure stability of the result. 
The classification results are evaluated in terms of Overall Accuracy (OA), Average Accuracy (AA) and Kappa statistic.
All our experiments have been conducted by using Matlab R2017b in a desktop PC equipped with an Intel Core i5 CPU (at 3.1GHz) and 8GB of RAM.

\section{Results and Discussions}
\begin{table*}[!t]
\setlength{\abovecaptionskip}{0pt}
\setlength{\belowcaptionskip}{0pt}
\caption{Classification results for the Pavia University Dataset. The Best Results For Each Column are Reported in Italic Bold. The Proposed \textbf{CDCL} Approach Outperforms All the Baseline Methods.}
\label{result_univ}
{\begin{tabularx}{1\textwidth}{c|c|XXXXXXXXX}
\toprule[2pt]
\multirow{2}{*}{Method} &\multirow{2}{*}{Metrics} &\multicolumn{9}{c}{Number of Source/Target Training Samples (per class)}\\
\cline{3-11}
{}&{}&\ \ \ \ 10/2&\ \ \ \ 20/2&\ \ \ \ 50/2&\ \ \ \ 10/3&\ \ \ \ 20/3&\ \ \ \ 50/3&\ \ \ \ 10/5&\ \ \ \ 20/5&\ \ \ \ 50/5 \\
\midrule[1pt]
\multirow{3}{*} {\emph{CCA}}&OA&$36.31/1.11$& $36.85/1.14$& $36.51/1.19$& $36.48/1.06$& $35.79/0.99$& $36.08/1.09$& $33.98/0.86$& $34.24/0.93$& $34.02/0.97$\\ 
{}&AA&$39.23/0.69$& $39.77/0.75$& $39.75/0.73$& $38.84/0.72$& $38.29/0.60$& $38.42/0.63$& $35.35/0.64$& $35.32/0.70$& $34.99/0.67$\\ 
{}&Kappa&$21.79/0.98$& $22.28/1.03$& $22.08/1.05$& $21.66/0.99$& $21.04/0.90$& $21.30/1.03$& $19.11/0.79$& $19.31/0.88$& $19.05/0.90$\\ 
\hline
\multirow{3}{*} {\emph{C-CCA}}&OA&$46.75/1.22$& $49.88/1.16$& $53.77/1.29$& $47.16/0.96$& $50.21/0.98$& $53.27/1.10$& $44.90/0.82$& $48.67/0.84$& $51.45/0.93$\\ 
{}&AA&$45.02/0.74$& $46.67/0.73$& $51.95/0.91$& $44.37/0.58$& $46.56/0.61$& $50.84/0.76$& $41.76/0.61$& $44.03/0.62$& $46.91/0.70$\\ 
{}&Kappa&$33.43/1.14$& $36.71/1.11$& $40.93/1.30$& $33.87/0.88$& $37.16/0.92$& $40.44/1.05$& $31.26/0.77$& $35.12/0.81$& $38.03/0.93$\\ 
\hline
\multirow{3}{*} {\emph{DAMA}}&OA&$48.50/0.90$& $43.70/1.23$& $41.53/1.03$& $51.31/1.04$& $46.97/1.31$& $46.03/1.17$& $54.35/0.81$& $51.66/1.03$& $52.50/0.96$\\ 
{}&AA&$47.53/0.57$& $51.21/0.72$& $55.49/0.71$& $49.62/0.70$& $52.81/0.65$& $57.04/0.62$& $51.91/0.65$& $53.86/0.68$& $59.99/0.47$\\ 
{}&Kappa&$37.28/0.83$& $33.88/1.10$& $32.52/0.94$& $39.98/1.00$& $36.95/1.18$& $36.67/1.07$& $42.70/0.81$& $40.87/0.99$& $42.73/0.89$\\  
\hline
\multirow{3}{*} {\emph{SSMA}}&OA&$53.55/1.30$& $52.77/1.44$& $47.53/1.15$& $59.72/1.32$& $55.99/1.22$& $53.42/1.03$& $62.78/1.08$& $60.84/1.05$& $58.12/0.95$\\ 
{}&AA&$59.01/1.01$& $59.90/1.02$& $59.34/0.76$& $63.81/0.86$& $61.61/0.75$& $61.50/0.59$& $67.47/0.71$& $65.56/0.65$& $64.26/0.58$\\ 
{}&Kappa&$42.50/1.31$& $42.20/1.45$& $37.67/1.05$& $49.13/1.35$& $45.28/1.19$& $43.19/1.01$& $52.83/1.11$& $50.75/1.08$& $48.14/0.96$\\
\hline
\multirow{3}{*} {\emph{KEMA}}&OA&$55.28/1.15$& $58.11/1.23$& $56.47/1.12$& $55.22/1.10$& $59.01/1.13$& $58.93/1.16$& $63.08/0.93$& $63.17/1.10$& $63.22/1.03$\\ 
{}&AA&$60.33/0.88$& $62.54/0.77$& $62.07/0.63$& $62.13/0.81$& $64.64/0.57$& $64.47/0.55$& $67.04/0.57$& $67.25/0.54$& $67.79/0.44$\\ 
{}&Kappa&$43.88/1.12$& $47.03/1.25$& $45.66/1.10$& $44.23/1.11$& $48.24/1.13$& $48.38/1.14$& $52.81/1.00$& $53.13/1.14$& $53.15/1.06$\\ 
\hline
\multirow{3}{*} {\emph{SHFA}}&OA&$56.59/1.12$& $56.21/1.10$& $55.13/1.14$& $58.36/0.99$& $58.28/0.98$& $57.77/1.00$& $61.35/0.89$& $61.10/0.87$& $61.11/0.90$\\ 
{}&AA&$59.99/0.70$& $59.05/0.73$& $58.66/0.76$& $62.11/0.60$& $62.08/0.62$& $61.95/0.61$& $65.06/0.56$& $64.89/0.55$& $64.82/0.56$\\ 
{}&Kappa&$44.57/1.09$& $44.24/1.08$& $43.27/1.10$& $46.80/0.98$& $46.76/0.98$& $46.30/0.99$& $50.53/0.91$& $50.25/0.88$& $50.28/0.92$\\ 
\hline
\multirow{3}{*} {\emph{CDLS}}&OA&$55.07/1.03$& $55.72/0.99$& $55.88/0.92$& $57.17/1.07$& $57.47/1.11$& $57.69/1.02$& $57.64/0.93$& $57.81/0.97$& $57.41/0.98$\\ 
{}&AA&$59.42/0.74$& $59.97/0.73$& $60.05/0.70$& $61.15/0.58$& $61.08/0.64$& $61.32/0.60$& $62.40/0.55$& $62.41/0.57$& $62.42/0.59$\\ 
{}&Kappa&$43.24/1.01$& $43.91/0.98$& $44.10/0.92$& $45.62/1.04$& $45.90/1.09$& $46.06/1.03$& $46.81/0.92$& $46.94/0.97$& $46.54/0.97$\\ 
\hline
\multirow{3}{*} {\emph{NA}}&OA&$58.88/1.14$& $58.88/1.14$& $58.88/1.14$& $62.50/1.00$& $62.50/1.00$& $62.50/1.00$& $67.40/0.84$& $67.40/0.84$& $67.40/0.84$\\ 
{}&AA&$64.16/0.62$& $64.16/0.62$& $64.16/0.62$& $66.16/0.62$& $66.16/0.62$& $66.16/0.62$& $70.09/0.55$& $70.09/0.55$& $70.09/0.55$\\ 
{}&Kappa&$47.99/1.16$& $47.99/1.16$& $47.99/1.16$& $51.87/1.04$& $51.87/1.04$& $51.87/1.04$& $57.70/0.92$& $57.70/0.92$& $57.70/0.92$\\ 
\hline
\multirow{3}{*} {\emph{LapSVM}}&OA&$55.42/1.10$& $55.42/1.10$& $55.42/1.10$& $59.11/1.01$& $59.11/1.01$& $59.11/1.01$& $60.80/0.84$& $60.80/0.84$& $60.80/0.84$\\
{}&AA&$56.05/0.70$& $56.05/0.70$& $56.05/0.70$& $58.59/0.68$& $58.59/0.68$& $58.59/0.68$& $61.12/0.63$& $61.12/0.63$& $61.12/0.63$\\ 
{}&Kappa&$43.81/1.04$& $43.81/1.04$& $43.81/1.04$& $47.73/1.01$& $47.73/1.01$& $47.73/1.01$& $49.83/0.85$& $49.83/0.85$& $49.83/0.85$\\
\hline
\multirow{3}{*} {\emph{ERW}}&OA&$70.05/1.41$& $70.05/1.41$& $70.05/1.41$& $72.93/1.25$& $72.93/1.25$& $72.93/1.25$& $83.12/1.10$& $83.12/1.10$& $83.12/1.10$\\
{}&AA&$77.29/1.00$& $77.29/1.00$& $77.29/1.00$& $78.58/0.80$& $78.58/0.80$& $78.58/0.80$& $85.11/0.81$& $85.11/0.81$& $\textbf{85.11}/0.81$\\
{}&Kappa&$59.79/1.55$& $59.79/1.55$& $59.79/1.55$& $64.32/1.43$& $64.32/1.43$& $64.32/1.43$& $77.09/1.39$& $77.09/1.39$& $77.09/1.39$\\
\hline
\multirow{3}{*} {\textbf{CDCL}}&OA&$\textbf{72.35}/1.62$& $\textbf{76.41}/1.50$& $\textbf{74.83}/1.55$& $\textbf{80.04}/1.29$& $\textbf{83.52}/1.11$& $\textbf{81.35}/1.11$& $\textbf{85.60}/1.08$& $\textbf{85.66}/1.11$& $\textbf{84.55}/1.08$\\
{}&AA&$\textbf{79.63}/0.93$& $\textbf{82.39}/0.84$& $\textbf{79.29}/0.87$& $\textbf{83.73}/0.74$& $\textbf{85.24}/0.82$& $\textbf{83.03}/0.72$& $\textbf{86.98}/0.71$& $\textbf{87.15}/0.70$& $84.51/0.78$\\
{}&Kappa&$\textbf{62.41}/1.91$& $\textbf{68.27}/1.73$& $\textbf{67.10}/1.75$& $\textbf{72.74}/1.58$& $\textbf{77.41}/1.39$& $\textbf{75.08}/1.31$& $\textbf{80.51}/1.31$& $\textbf{80.76}/1.35$& $\textbf{79.39}/1.31$\\
\bottomrule[2pt]
\end{tabularx}}
\end{table*}
\begin{figure*}[!t]
\setlength{\abovecaptionskip}{-5pt}
\setlength{\belowcaptionskip}{-20pt}
\centering
\includegraphics[width=1\textwidth]{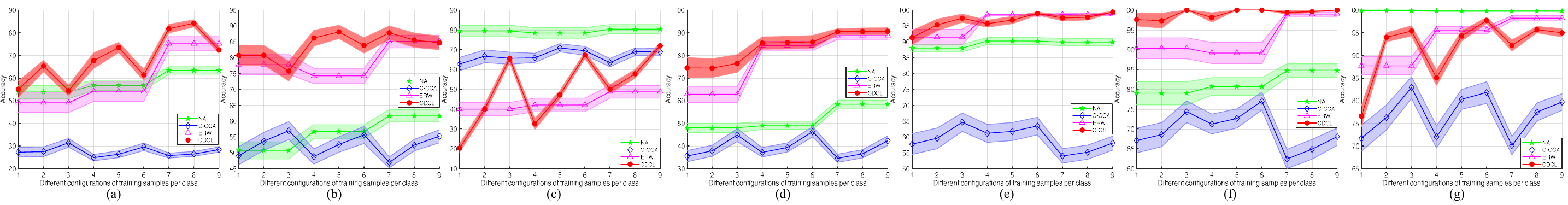}
\caption{Individual accuracies of different classes (a-g) obtained by \emph{NA}, \emph{C-CCA}, \emph{ERW} and \textbf{CDCL} methods on the Pavia University dataset. Note that the different configurations of training samples correspond to the settings in Table \ref{result_univ}.}
\label{Acc_univ}
\end{figure*}

\subsection{Results of \emph{Univ/Center} Case}
To illustrate the effectiveness of the proposed \textbf{CDCL} on the whole HSI, an experiment is performed with the setting (TR\_T and TR\_S) in Table \ref{Setup} and all unlabeled samples in Pavia Center HSI as TE\_T. Fig. \ref{CenterMap}(a)-(k) show the classification results obtained by different methods, including \emph{CCA}, \emph{C-CCA}, \emph{DAMA}, \emph{SSMA}, \emph{KEMA}, \emph{SHFA}, \emph{CDLS}, \emph{NA}, \emph{LapSVM}, \emph{ERW} and the proposed \textbf{CDCL} methods. Fig. \ref{CenterMap}(l) represents the corresponding ground truth. From this figure, it can be seen that the \textbf{CDCL} method can effectively remove the noise in the \emph{NA} and \emph{ERW} classification results. Furthermore, the \textbf{CDCL} method obtained the highest OA = 91.03\%. Table \ref{result_city} reports the results of different methods in terms of individual class accuracies, the mean and standard variance of OA, AA, and Kappa statistics using the setting in Table \ref{Setup}. The following observations can be done:\\
\begin{table*}[!t]
\setlength{\abovecaptionskip}{0pt}
\setlength{\belowcaptionskip}{0pt}
\caption{Classification results for the Salinas Dataset. The Best Results For Each Row are Reported in Italic Bold. The Proposed \textbf{CDCL} Approach Outperforms All the Baseline Methods.}
\label{result_salinas}
{\begin{tabularx}{1\textwidth}{c|X|X|X|X|X|X|X|X|X|X|X}
\toprule[2pt]
\multirow{2}{*}{Class} &\multicolumn{11}{c}{Methods}\\
\cline{2-12}
{}  & \emph{CCA} & \emph{C-CCA}  & \emph{DAMA} & \emph{SSMA} & \emph{KEMA} & \emph{SHFA} & \emph{CDLS} & \emph{NA} & \emph{LapSVM} & \emph{ERW} & \textbf{CDCL}\\
\midrule[1pt]
\emph{Weeds\_1}& $48.05/2.7$& $97.41/0.6$& $98.44/0.5$& $98.73/0.5$& $95.90/0.9$& $96.88/0.7$& $89.61/2.3$& $95.76/0.9$& $95.85/0.8$& $97.22/1.1$& $\textbf{100.0}/0.0$\\ 
\emph{Weeds\_2}& $45.25/2.8$& $94.45/1.0$& $98.56/0.5$& $98.40/0.5$& $89.68/1.8$& $94.88/0.7$& $74.72/3.3$& $84.91/1.6$& $91.12/1.1$& $91.09/1.8$& $\textbf{100.0}/0.0$\\ 
\emph{Fallow}& $23.50/2.0$& $77.10/1.8$& $82.00/2.8$& $76.50/2.2$& $75.50/2.6$& $53.95/1.7$& $43.55/3.8$& $67.40/2.4$& $57.15/2.3$& $\textbf{100.0}/0.0$& $\textbf{100.0}/0.0$\\ 
\emph{Fallow\_r}& $41.64/2.8$& $95.57/0.9$& $99.43/0.2$& $99.36/0.2$& $97.86/0.5$& $99.36/0.2$& $76.21/5.7$& $98.86/0.3$& $99.07/0.3$& $69.21/3.3$& $\textbf{99.79}/0.1$\\ 
\emph{Fallow\_s}& $32.67/2.6$& $76.41/2.6$& $78.00/3.0$& $87.81/2.6$& $91.41/1.7$& $98.41/0.4$& $24.19/5.8$& $91.63/1.5$& $95.67/1.2$& $96.74/1.5$& $\textbf{99.52}/0.1$\\ 
\emph{Stubble}& $54.02/2.6$& $99.08/0.1$& $99.35/0.1$& $99.40/0.1$& $97.10/0.6$& $99.47/0.1$& $93.03/2.4$& $96.78/0.4$& $98.35/0.3$& $97.22/0.4$& $\textbf{99.90}/0.0$\\ 
\emph{Celery}& $46.28/2.7$& $98.81/0.3$& $98.53/0.2$& $98.19/0.5$& $95.89/1.0$& $97.56/0.6$& $30.69/5.9$& $93.00/1.5$& $93.33/1.3$& $90.78/1.9$& $\textbf{100.0}/0.0$\\ 
\emph{Graphes\_u}& $28.61/2.2$& $60.63/2.7$& $18.85/2.7$& $40.15/3.3$& $57.71/3.4$& $49.98/2.8$& $42.05/2.9$& $46.23/2.7$& $52.35/2.8$& $48.97/4.1$& $\textbf{70.16}/2.0$\\ 
\emph{Soil\_v}& $33.68/2.2$& $92.48/1.1$& $81.25/3.5$& $85.74/2.2$& $94.72/1.0$& $97.04/0.2$& $21.46/5.5$& $90.98/1.2$& $93.92/0.6$& $\textbf{100.0}/0.0$& $\textbf{100.0}/0.0$\\ 
\emph{Corn\_s}& $33.45/2.3$& $75.18/1.9$& $57.76/2.4$& $60.30/2.3$& $67.33/2.2$& $72.45/2.5$& $58.33/2.4$& $59.73/2.1$& $66.39/2.4$& $88.24/2.6$& $\textbf{97.70}/0.3$\\ 
\emph{Lettuce\_4wk}& $45.55/3.0$& $92.91/1.2$& $88.64/1.6$& $90.27/1.0$& $84.45/2.3$& $86.64/2.0$& $47.36/5.9$& $84.82/2.2$& $89.36/1.9$& $\textbf{99.91}/0.1$& $99.36/0.2$\\ 
\emph{Lettuce\_5wk}& $25.49/2.3$& $90.56/1.3$& $99.69/0.1$& $98.87/0.5$& $84.15/2.0$& $97.69/1.1$& $48.26/4.0$& $87.54/2.0$& $86.00/2.8$& $\textbf{100.0}/0.0$& $\textbf{100.0}/0.0$\\ 
\emph{Lettuce\_6wk}& $28.40/2.4$& $86.50/1.9$& $98.40/0.4$& $98.30/0.6$& $97.50/0.5$& $98.30/0.4$& $93.80/1.0$& $96.80/1.1$& $96.80/1.0$& $95.20/2.4$& $\textbf{99.40}/0.3$\\ 
\emph{Lettuce\_7wk}& $35.45/2.9$& $78.09/3.2$& $86.82/2.0$& $82.45/2.1$& $86.64/2.6$& $87.36/1.7$& $62.36/4.1$& $80.36/3.3$& $79.73/2.8$& $77.82/2.7$& $\textbf{99.18}/0.2$\\ 
\emph{Vinyard\_u}& $30.49/2.0$& $60.25/2.5$& $\textbf{88.97}/2.3$& $75.14/2.8$& $60.45/2.9$& $60.84/2.5$& $54.59/3.6$& $60.59/2.7$& $57.11/2.7$& $67.19/4.5$& $84.62/1.4$\\ 
\emph{Vinyard\_v}& $42.16/2.5$& $92.05/1.5$& $83.14/1.4$& $85.57/1.2$& $89.35/1.8$& $75.46/2.6$& $51.73/2.8$& $66.38/2.6$& $67.57/2.2$& $51.95/3.8$& $\textbf{99.78}/0.1$\\ 
\midrule[1pt]
OA&$35.80/0.9$& $80.01/0.4$& $73.88/0.7$& $77.37/0.5$& $79.12/0.5$& $78.51/0.5$& $51.42/2.0$& $74.28/0.6$& $76.28/0.5$& $79.72/0.7$& $\textbf{91.55}/0.3$\\ 
AA&$50.35/0.6$& $84.98/0.4$& $86.73/0.4$& $87.16/0.3$& $85.36/0.4$& $85.86/0.4$& $51.00/2.4$& $79.54/0.5$& $81.42/0.5$& $87.18/0.5$& $\textbf{96.57}/0.1$\\ 
Kappa&$30.43/0.1$& $77.84/0.0$& $71.36/0.1$& $75.04/0.1$& $76.86/0.1$& $76.22/0.1$& $47.18/0.2$& $71.60/0.1$& $73.78/0.1$& $77.55/0.1$& $\textbf{90.64}/0.0$\\ 
\bottomrule[2pt]
\end{tabularx}}
\end{table*}
$\bullet$ The \textbf{CDCL}  method gives the highest classification accuracies for \emph{``Baresoil''}, \emph{``Bricks''} and \emph{``Bitumen''} classes. Moreover, the \textbf{CDCL} method also shows the best performance in terms of OA = 83.24\%, AA = 82.29\%, and Kappa = 80.00\%. \\
$\bullet$ The results of \emph{KEMA} and \emph{SHFA} are comparable and better than results of other HDA methods, whereas the \emph{CCA} method performs worst due to the fact that only part of labeled samples are used.\\
$\bullet$ The \emph{NA} method outperforms \emph{LapSVM} and \emph{ERW} methods, and even all the baseline HDA methods. It can be concluded that the knowledge of Pavia University data can hardly be well transferred to the Center data with limited target labeled samples. In addition, both \textbf{CDCL} and \emph{ERW} perform worse than \emph{NA} method on the \emph{``Meadows''} and \emph{``Shadows''} classes, confirming the relation between \emph{ERW} and \textbf{CDCL} methods.

\begin{figure}
\setlength{\abovecaptionskip}{-5pt}
\setlength{\belowcaptionskip}{-20pt}
\centering
\includegraphics[width=0.5\textwidth]{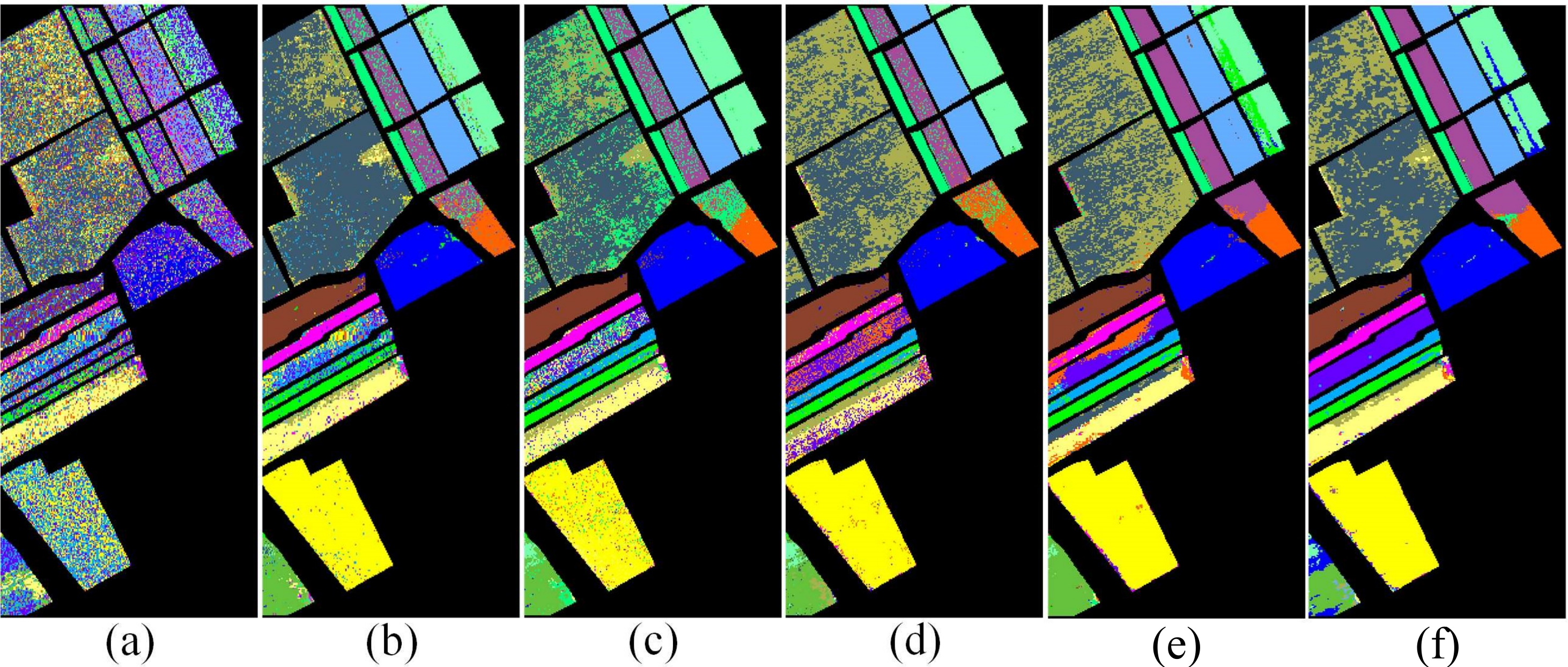}
\includegraphics[width=0.5\textwidth]{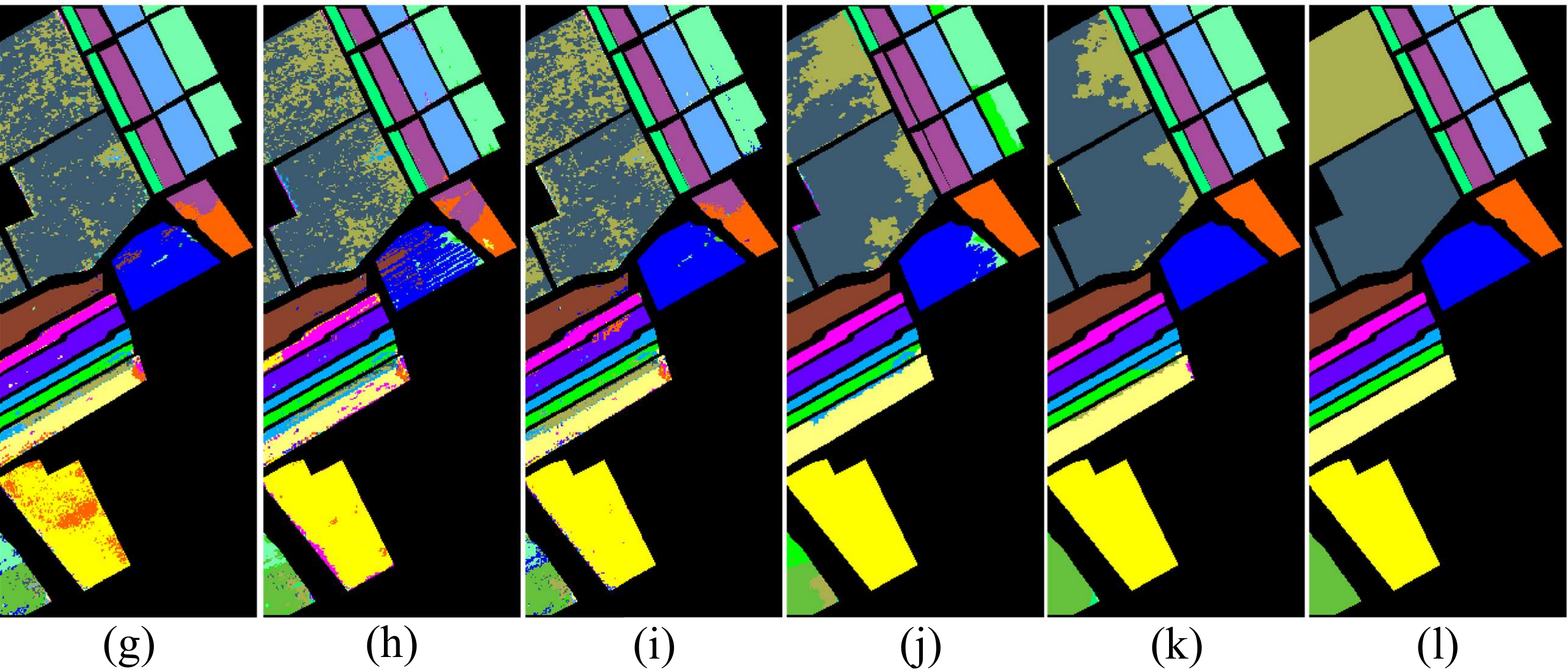}
\caption{Classification map of Salinas Image by (a)  \emph{CCA} (OA=36.07\%), (b) \emph{C-CCA} (OA=76.19\%),  (c) \emph{DAMA} (OA=73.90\%), (d) \emph{SSMA} (OA=77.02\%),  (e) \emph{KEMA} (OA=74.27\%), (f) \emph{SHFA} (OA=81.48\%),  (g) \emph{CDLS} (OA=79.27\%), (h) \emph{NA} (OA=78.17\%),  (i) \emph{LapSVM} (OA=81.36\%), (j) \emph{ERW} (OA=83.40\%),  (k) \emph{CDCL} (OA=88.13\%) methods and (l) denotes the corresponding ground truth.}
\label{SalinasMap}
\end{figure}
\subsection{Results of \emph{Center/Univ} Case}
Table \ref{result_univ} illustrates the OAs, AAs, Kappa statistics and the corresponding standard errors obtained by the proposed \textbf{CDCL} method and the baseline methods for the \emph{Center/Univ} case.
The experiments are performed with different numbers of source and target training samples illustrated in Table \ref{Setup}.
The following observations can be easily drawn:\\
$\bullet$ When increasing the number of labeled source and target samples, the mean OAs, AAs and Kappa statistics of most methods increase as expected. The increasing trend of mean OAs with more target training samples confirms that 50 trials are enough for achieving stable results. Moreover, the standard errors of OAs, AAs and Kappa statistics for smaller numbers of labeled samples appear to be higher.\\
$\bullet$ The \textbf{CDCL}  method gives the highest classification accuracies with different numbers of training samples. To be specific, the mean OAs of \emph{NA}, \emph{ERW} and \textbf{CDCL} methods are in the range of 58.88\%-67.4\%, 70.05\%-83.12\% and 72.35\%-85.66\%, respectively. Further, when only 10 per class source labeled samples are used for training, the \textbf{CDCL} method yields 2.30\%, 7.11\% and 2.48\% higher mean OAs than \emph{ERW} with 2, 3 and 5 target samples per class. 

Fig. \ref{Acc_univ} reports individual class accuracies for the \emph{Center/Univ} case obtained by \emph{C-CCA}, \emph{NA}, \emph{ERW} and \textbf{CDCL} methods using different numbers of labeled samples, assessed by the mean OAs (main curves) and their standard errors (shaded area for each curve). The classification accuracies of 7 classes (\emph{``asphalt''}, \emph{``meadows''}, \emph{``trees''}, \emph{``baresoil''}, \emph{``bricks''}, \emph{``bitumen''}, \emph{``shadows''}) are shown in Fig. \ref{Acc_univ}(a-g), respectively. Note that the abscissas represent the different settings of training number in Table \ref{result_univ}. The \textbf{CDCL} method outperforms \emph{C-CCA}, \emph{NA} and \emph{ERW} methods on \emph{``asphalt''} (a), \emph{``meadows''} (b), \emph{``baresoil''} (d) and \emph{``bitumen''} (f) classes, and shows comparable accuracy on \emph{``bricks''} class (e) with the \emph{ERW} method, yielding a better overall classification accuracy. Further, the \emph{ERW} method performs worse than \emph{NA} on \emph{``trees''} (c) and \emph{``shadows''} (g) classes, resulting in low accuracies of \textbf{CDCL} method on these two classes.
\subsection{Results of \emph{Salinas} Case}
\begin{table*}[!t]
\setlength{\abovecaptionskip}{0pt}
\setlength{\belowcaptionskip}{0pt}
\caption{Classification results for the Indian Pines Dataset. The Best Results For Each Column are Reported in Italic Bold. The Proposed \textbf{CDCL} Approach Outperform All the Baseline Methods.}
\label{result_indian}
{\begin{tabularx}{1\textwidth}{c|c|XXXXXXXXX}
\toprule[2pt]
\multirow{2}{*}{Method} &\multirow{2}{*}{Metrics} &\multicolumn{9}{c}{Number of Source/Target Training Samples (per class)}\\
\cline{3-11}
{}&{}&\ \ \ \ 5/2&\ \ \ \ 10/2&\ \ \ \ 15/2&\ \ \ \ 5/3&\ \ \ \ 10/3&\ \ \ \ 15/3&\ \ \ \ 5/5&\ \ \ \ 10/5&\ \ \ \ 15/5 \\
\midrule[1pt]
\multirow{3}{*} {\emph{CCA}}&OA&$22.24/0.59$& $21.99/0.61$& $21.81/0.62$& $19.74/0.50$& $19.51/0.49$& $19.56/0.53$& $18.53/0.35$& $18.40/0.36$& $17.88/0.37$\\ 
{}&AA&$36.30/0.88$& $35.65/0.90$& $36.02/0.74$& $36.69/0.86$& $36.30/0.86$& $36.50/0.85$& $28.65/0.66$& $27.54/0.72$& $27.32/0.59$\\ 
{}&Kappa&$16.54/0.62$& $16.29/0.64$& $16.14/0.66$& $13.78/0.53$& $13.54/0.52$& $13.61/0.56$& $12.79/0.38$& $12.72/0.40$& $12.25/0.38$\\ 
\hline
\multirow{3}{*} {\emph{C-CCA}}&OA&$38.03/0.59$& $39.95/0.57$& $40.43/0.64$& $38.76/0.56$& $40.25/0.51$& $42.00/0.57$& $40.20/0.51$& $42.27/0.45$& $43.19/0.48$\\ 
{}&AA&$44.16/0.70$& $45.37/0.68$& $45.10/0.67$& $45.00/0.61$& $45.17/0.56$& $46.30/0.56$& $47.15/0.51$& $47.81/0.45$& $47.98/0.47$\\ 
{}&Kappa&$33.58/0.62$& $35.69/0.59$& $36.22/0.67$& $34.38/0.59$& $35.97/0.55$& $37.89/0.61$& $35.93/0.54$& $38.21/0.47$& $39.26/0.51$\\
\hline
\multirow{3}{*} {\emph{DAMA}}&OA&$45.40/0.62$& $46.14/0.53$& $46.13/0.52$& $49.47/0.54$& $50.99/0.48$& $51.02/0.48$& $54.07/0.46$& $55.19/0.44$& $55.82/0.52$\\ 
{}&AA&$51.36/0.76$& $51.74/0.64$& $51.99/0.62$& $54.87/0.68$& $55.69/0.59$& $55.66/0.60$& $58.12/0.45$& $59.16/0.51$& $59.65/0.56$\\ 
{}&Kappa&$41.29/0.62$& $41.95/0.55$& $42.07/0.53$& $45.75/0.55$& $47.19/0.49$& $47.35/0.51$& $50.73/0.48$& $51.88/0.46$& $52.58/0.55$\\
\hline
\multirow{3}{*} {\emph{SSMA}}&OA&$44.51/0.49$& $45.77/0.46$& $46.79/0.44$& $49.11/0.54$& $50.01/0.54$& $50.49/0.45$& $53.46/0.45$& $54.55/0.51$& $55.71/0.41$\\ 
{}&AA&$49.62/0.65$& $50.48/0.64$& $51.73/0.56$& $53.95/0.61$& $54.48/0.65$& $55.07/0.61$& $58.06/0.54$& $58.51/0.57$& $59.33/0.49$\\ 
{}&Kappa&$40.57/0.50$& $41.83/0.47$& $42.86/0.46$& $45.50/0.56$& $46.35/0.56$& $46.85/0.47$& $50.16/0.48$& $51.23/0.54$& $52.43/0.44$\\ 
\hline
\multirow{3}{*} {\emph{KEMA}}&OA&$43.98/0.56$& $44.77/0.52$& $44.98/0.57$& $45.32/0.54$& $46.42/0.52$& $46.89/0.49$& $47.58/0.49$& $49.15/0.35$& $49.36/0.36$\\ 
{}&AA&$48.63/0.62$& $49.35/0.49$& $50.28/0.56$& $50.05/0.52$& $51.64/0.56$& $52.11/0.53$& $52.82/0.57$& $53.75/0.42$& $55.20/0.41$\\ 
{}&Kappa&$40.02/0.58$& $40.85/0.54$& $41.08/0.59$& $41.49/0.58$& $42.68/0.55$& $43.16/0.52$& $43.92/0.52$& $45.61/0.37$& $45.84/0.39$\\ 
\hline
\multirow{3}{*} {\emph{SHFA}}&OA&$48.72/0.52$& $48.75/0.51$& $48.83/0.53$& $52.20/0.58$& $52.31/0.60$& $52.27/0.57$& $58.26/0.46$& $58.25/0.46$& $58.24/0.44$\\ 
{}&AA&$52.44/0.50$& $52.50/0.51$& $52.53/0.51$& $55.43/0.62$& $55.54/0.61$& $55.47/0.59$& $61.46/0.46$& $61.46/0.46$& $61.47/0.44$\\ 
{}&Kappa&$45.07/0.54$& $45.10/0.53$& $45.19/0.56$& $48.82/0.62$& $48.94/0.63$& $48.89/0.60$& $55.33/0.48$& $55.31/0.48$& $55.30/0.46$\\ 
\hline
\multirow{3}{*} {\emph{CDLS}}&OA&$47.78/0.61$& $47.94/0.58$& $48.09/0.58$& $50.05/0.59$& $50.32/0.53$& $50.31/0.57$& $54.91/0.44$& $54.66/0.50$& $54.39/0.55$\\ 
{}&AA&$51.49/0.61$& $51.60/0.57$& $52.04/0.58$& $53.91/0.71$& $53.97/0.61$& $53.92/0.63$& $58.42/0.49$& $58.13/0.50$& $58.27/0.49$\\ 
{}&Kappa&$44.13/0.65$& $44.31/0.61$& $44.47/0.61$& $46.60/0.62$& $46.88/0.56$& $46.84/0.60$& $51.73/0.47$& $51.50/0.52$& $51.24/0.57$\\ 
\hline
\multirow{3}{*} {\emph{NA}}&OA&$48.48/0.57$& $48.48/0.57$& $48.48/0.57$& $53.17/0.58$& $53.17/0.58$& $53.17/0.58$& $58.35/0.48$& $58.35/0.48$& $58.35/0.48$\\ 
{}&AA&$51.91/0.56$& $51.91/0.56$& $51.91/0.56$& $56.80/0.58$& $56.80/0.58$& $56.80/0.58$& $61.53/0.46$& $61.53/0.46$& $61.53/0.46$\\ 
{}&Kappa&$44.88/0.60$& $44.88/0.60$& $44.88/0.60$& $49.90/0.61$& $49.90/0.61$& $49.90/0.61$& $55.44/0.50$& $55.44/0.50$& $55.44/0.50$\\ 
\hline
\multirow{3}{*} {\emph{LapSVM}}&OA&$47.33/0.59$& $47.33/0.59$& $47.33/0.59$& $49.87/0.48$& $49.87/0.48$& $49.87/0.48$& $54.96/0.48$& $54.96/0.48$& $54.96/0.48$\\ 
{}&AA&$51.75/0.57$& $51.75/0.57$& $51.75/0.57$& $54.57/0.52$& $54.57/0.52$& $54.57/0.52$& $58.98/0.43$& $58.98/0.43$& $58.98/0.43$\\ 
{}&Kappa&$43.67/0.61$& $43.67/0.61$& $43.67/0.61$& $46.38/0.51$& $46.38/0.51$& $46.38/0.51$& $51.82/0.50$& $51.82/0.50$& $51.82/0.50$\\ 
\hline
\multirow{3}{*} {\emph{ERW}}&OA&$61.06/0.73$& $61.06/0.73$& $61.06/0.73$& $70.43/0.67$& $70.43/0.67$& $70.43/0.67$& $81.38/0.47$& $81.38/0.47$& $81.38/0.47$\\ 
{}&AA&$72.86/0.59$& $72.86/0.59$& $72.86/0.59$& $78.46/0.50$& $78.46/0.50$& $78.46/0.50$& $84.57/0.37$& $84.57/0.37$& $84.57/0.37$\\ 
{}&Kappa&$58.30/0.76$& $58.30/0.76$& $58.30/0.76$& $68.39/0.71$& $68.39/0.71$& $68.39/0.71$& $80.07/0.50$& $80.07/0.50$& $80.07/0.50$\\ 
\hline
\multirow{3}{*} {\emph{CDCL}}&OA&$\textbf{74.92}/0.64$& $\textbf{77.78}/0.57$& $\textbf{78.48}/0.61$& $\textbf{79.81}/0.56$& $\textbf{81.80}/0.55$& $\textbf{82.75}/0.46$& $\textbf{86.01}/0.37$& $\textbf{86.24}/0.35$& $\textbf{87.06}/0.35$\\ 
{}&AA&$\textbf{81.45}/0.50$& $\textbf{82.62}/0.50$& $\textbf{82.94}/0.48$& $\textbf{84.41}/0.45$& $\textbf{85.37}/0.46$& $\textbf{86.09}/0.37$& $\textbf{88.68}/0.30$& $\textbf{88.91}/0.26$& $\textbf{89.43}/0.26$\\ 
{}&Kappa&$\textbf{73.13}/0.68$& $\textbf{76.23}/0.60$& $\textbf{77.00}/0.64$& $\textbf{78.38}/0.60$& $\textbf{80.53}/0.59$& $\textbf{81.54}/0.49$& $\textbf{85.01}/0.40$& $\textbf{85.26}/0.37$& $\textbf{86.15}/0.37$\\ 
\bottomrule[2pt]
\end{tabularx}}
\end{table*}
\begin{figure*}[!t]
\setlength{\abovecaptionskip}{-5pt}
\setlength{\belowcaptionskip}{-20pt}
\centering
\includegraphics[width=1\textwidth]{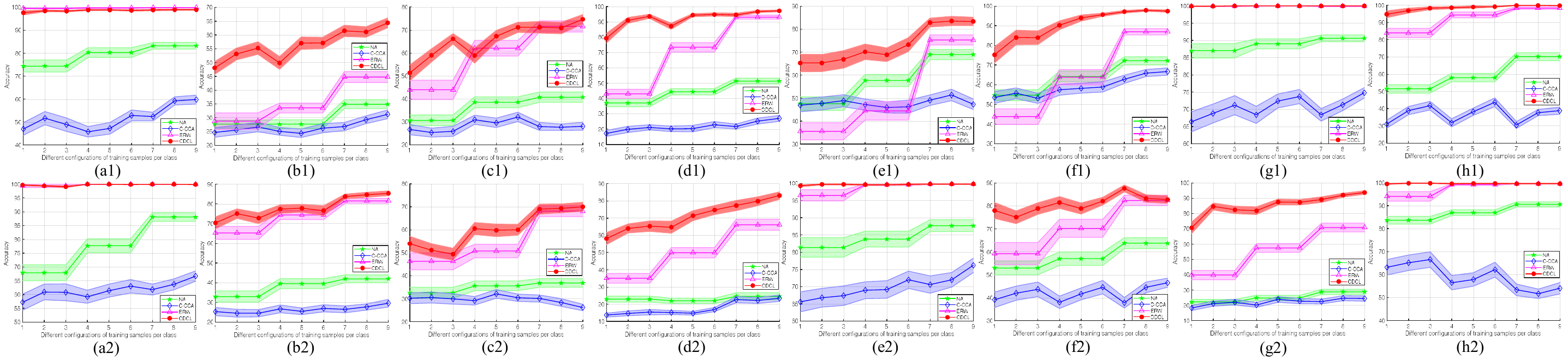}
\caption{Individual accuracies of different classes (a-g) obtained by \emph{NA}, \emph{C-CCA}, \emph{ERW} and \textbf{CDCL} methods on the Indian Pines dataset. Note that the different configurations of training samples correspond to the settings in Table \ref{result_indian}.}
\label{Acc_indian}
\end{figure*}
Similarly to the experiment of Pavia Center dataset, an experiment is firstly performed with the setting (TR\_T and TR\_S) in Table \ref{Setup} and the whole Salinas image as TE\_T. Fig. \ref{SalinasMap}(a)-(k) show the classification results obtained by all methods. Fig. \ref{SalinasMap}(l) shows the ground truth of the Salinas image. It is clear that the proposed \textbf{CDCL} method can effectively remove the noise in the \emph{NA} classification result. Furthermore, the \textbf{CDCL} method obtained the highest OA = 88.13\%. Table \ref{result_salinas} reports the results of different methods in terms of individual class accuracies, the mean and standard variance of OA, AA, and Kappa using the setting in Table \ref{Setup}.
The following observations can be easily obtained:\\
$\bullet$ The \textbf{CDCL} method obtained the highest classification accuracies except \emph{``Lettuce\_4wk''} and \emph{``Vinyard\_u''} class. Moreover, the \textbf{CDCL} method also shows the best performance in terms of OA = 91.55\%, AA = 96.57\%, and Kappa = 90.64\%. Compared with \textbf{CDCL}, OAs obtained by the \emph{NA} and \emph{ERW} methods are in average 17.27\% and 11.83\% lower, respectively. \\
$\bullet$ Among the HDA baseline methods, only the \emph{CCA}, \emph{DAMA} and \emph{CDLS} methods perform worse than \emph{NA}, yielding $\sim$38\%, $\sim$1\% and $\sim$22.86\% lower mean OAs, respectively. Similar to the result in \emph{Univ/Center} case, mean OA achieved by the \emph{CCA} method is more than 30\% lower compared with the \emph{NA} method. However, the observation that the \emph{C-CCA}, \emph{SSMA}, \emph{KEMA} and \emph{SHFA} methods outperform \emph{NA} method demonstrates the effectiveness of DA techniques. \\
$\bullet$ The \emph{LapSVM} and \emph{ERW} methods outperform \emph{NA} method, with $\sim$2.0\% and $\sim$5.5\% higher mean OAs. Furthermore, \textbf{CDCL} and \emph{ERW} methods show different trends on the \emph{``Fallow\_r''}, \emph{``Celery''}, \emph{``Lettuce\_6wk''}, \emph{``Lettuce\_7wk''} and  \emph{``Vinyard\_v''} classes. To be specific, the \emph{ERW} method performs worse than the \emph{NA} method, whereas the \textbf{CDCL} method performs better than the \emph{ERW} on the five classes. The phenomenon reveals that the cross-domain learning shows a direct improvement on the classification performance of the five classes. 
\subsection{Results of \emph{Indian} Case}
Table \ref{result_indian} illustrates the OAs, AAs, Kappa statistics and the corresponding standard errors obtained by all methods for \emph{Indian} case.
The experiments are performed 50 times with different numbers of source and target training samples illustrated in Table \ref{Setup}.
The following observations can be easily drawn:\\
$\bullet$ As expected, most methods perform better with the availability of more labeled samples in both domains. Compared with the \emph{NA} method, most HDA methods perform worse in most cases. However, \emph{SHFA} performs same as \emph{NA} when 5 labeled target samples per class  are used. It can be deduced that performance of \emph{SHFA} are affected by the number of target training samples.\\
$\bullet$ The \textbf{CDCL}  method gives the highest classification accuracies with different numbers of training samples. To be specific, the mean OAs of \emph{NA}, \emph{ERW} and \textbf{CDCL} methods are in the range of 48.48\%-58.35\%, 61.06\%-81.38\% and 74.92\%-87.06\%, respectively. Further, compared with \emph{NA}, the \textbf{CDCL} method yields 26.44\%-30\% higher mean OAs, depending on the number of training samples.

Fig. \ref{Acc_indian} reports individual class accuracies obtained by the \emph{C-CCA}, \emph{NA}, \emph{ERW} and \textbf{CDCL} methods using different numbers of labeled samples, assessed by the mean OAs (main curves) and their standard errors (shaded area for each curve). The classification accuracies of 16 classes (\emph{``Alfalfa''}, \emph{``Corn\_n''}, \emph{``Corn\_m''}, \emph{``Corn''}, \emph{``Grass-pasture''}, \emph{``Grass-trees''}, \emph{``Grass-pasture\_m''}, \emph{``Hay\_w''}, \emph{``Oats''}, \emph{``Soybean\_n''}, \emph{``Soybean\_m''}, \emph{``Soybean\_c''}, \emph{``Wheat''}, \emph{``Woods''}, \emph{``Buildings-Grass''}, \emph{``Stone-Steel''}) are shown in Fig. \ref{Acc_indian}(a1-h1) and (a2-h2), respectively. Similarly to Fig \ref{Acc_univ}, the abscissas represent the different settings of the number of training samples in Table \ref{result_indian}. The \textbf{CDCL} method outperforms the \emph{C-CCA}, \emph{NA} and \emph{ERW} methods on most classes.

\subsection{Adaptation Analysis}
\begin{figure}
\setlength{\abovecaptionskip}{-5pt}
\setlength{\belowcaptionskip}{-20pt}
\centering
\includegraphics[width=0.5\textwidth]{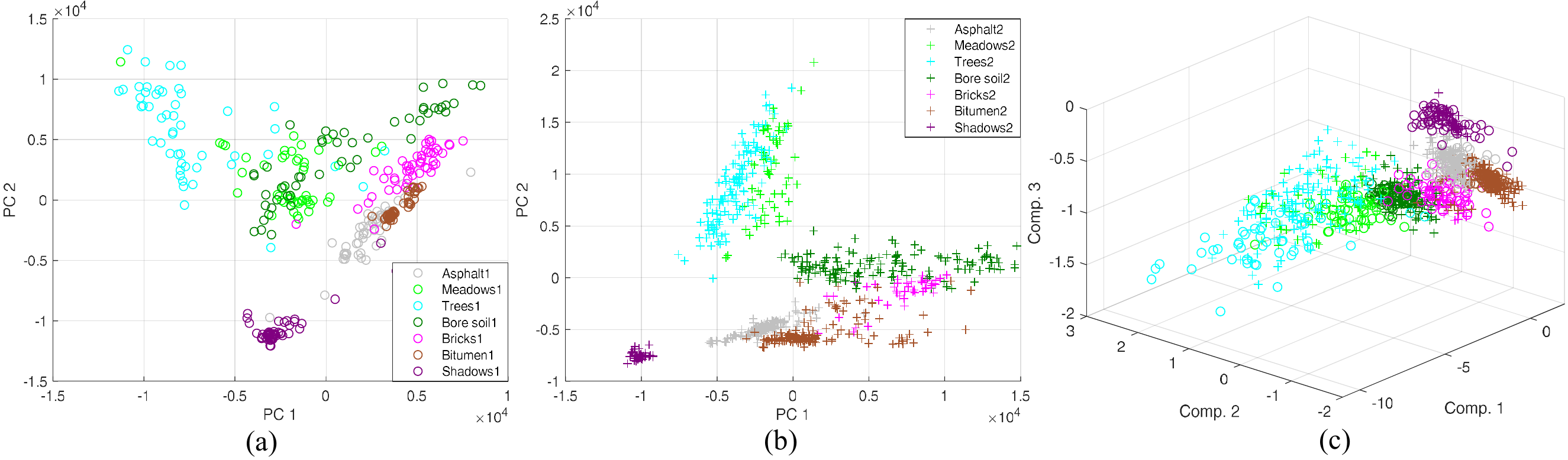}
\includegraphics[width=0.5\textwidth]{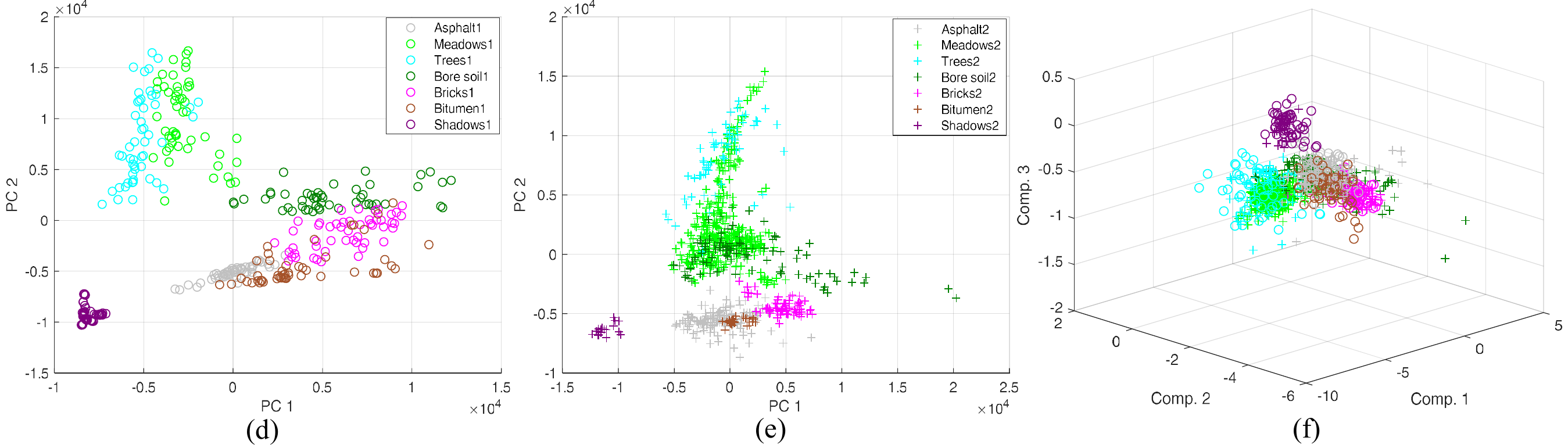}
\includegraphics[width=0.5\textwidth]{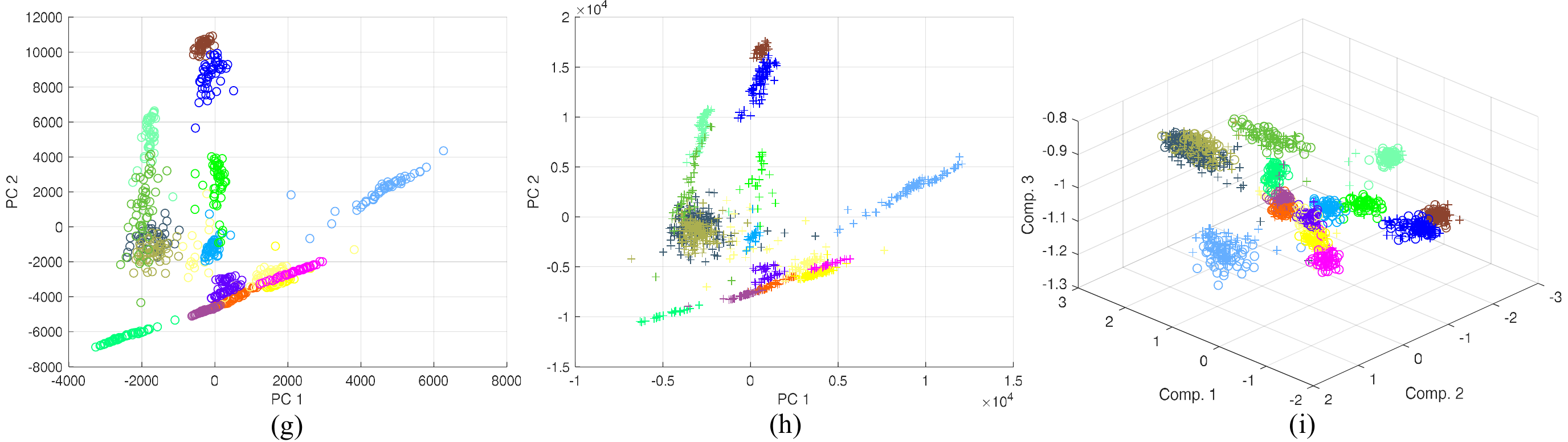}
\includegraphics[width=0.5\textwidth]{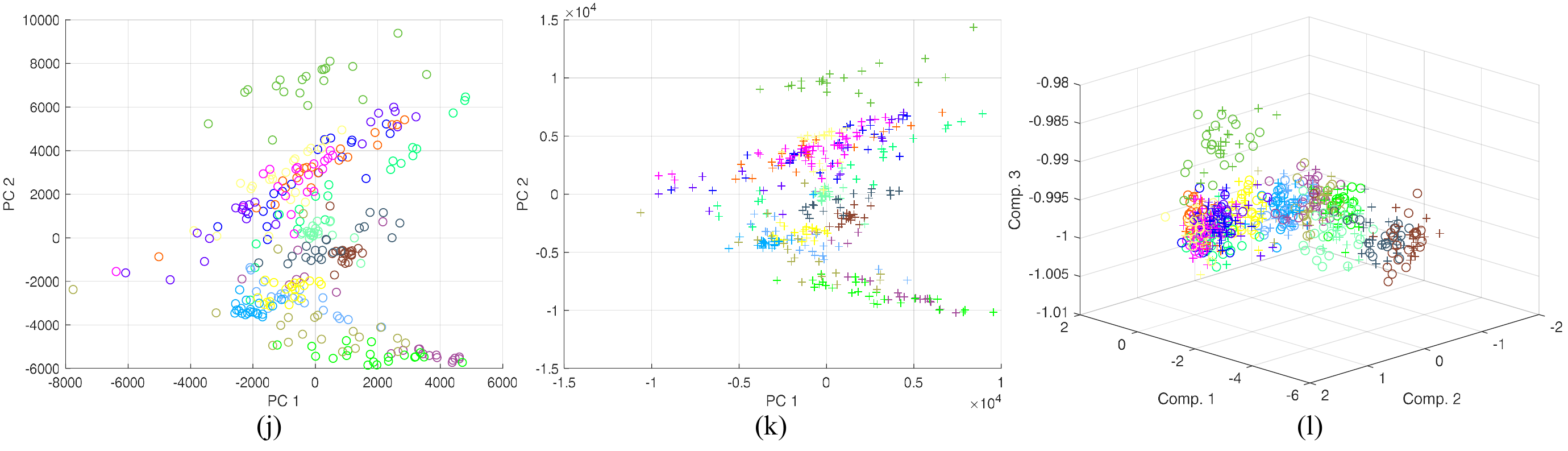}
\caption{Illustrative examples of aligned samples from different domains for four cases. Each row corresponds to one case (up to down: \emph{Univ/Center}, \emph{Center/Univ}, \emph{Salinas}, \emph{Indian} cases). Each column corresponds to one domain (left to right: source, target and correlated subspace). Note that the classes in the last two cases are ignored for better illustration. (Best viewed in color).}
\label{Scatter}
\end{figure}

Fig. \ref{Scatter} shows examples of aligned samples for the four cases.
As we can see from Fig. \ref{Scatter}, samples from both domains are well aligned in the correlated subspace for the Pavia dataset. Further, although Salinas and Indian Pines are totally different HSIs, HDA can also be achieved with the availability of labeled samples from both domains.
It is notable that the raw spectral features of the Salinas data have better-separated cluster structures than in Indian Pines data, resulting in better classification performance in the \emph{Salinas} case than the \emph{Indian} case.
\subsection{Computational Cost} 
\begin{table}[!t]
\setlength{\abovecaptionskip}{0pt}
\setlength{\belowcaptionskip}{0pt}
\caption{Computational Time (in seconds) of 
Different  Methods on Four Cases Using Different Numbers of Training Samples. The Largest Costs for Each Column are Reported in Italic Bold.}
\label{Cost}
{\begin{tabular}{@{}l@{}c@{}ccccc@{}}
\toprule[1.5pt]
\multirow{2}{*}{Methods}&\emph{Univ/Center}& \multicolumn{2}{c}{\emph{Center/Univ}}&\emph{Salinas}& \multicolumn{2}{c}{\emph{Indian} } \\
\cline{3-4}
\cline{6-7}
{}&50/2&10/2&50/5&50/2&5/2&15/5\\
\hline
{CCA \cite{Hardoon2004}}&$0.001$& $0.001$& $0.002$& $0.004$& $0.002$& $0.003$\\
{C-CCA \cite{Rasiwasia2014}}&$0.019$& $0.009$& $0.019$& $0.018$& $0.013$& $0.019$\\
{DAMA \cite{Wang2011}}&$1.357$& $2.361$& $2.407$& $0.370$& $0.094$& $0.125$\\
{SSMA \cite{Tuia2014}}&$1.678$& $2.863$& $2.955$& $0.400$& $0.106$& $0.140$\\
{KEMA \cite{Tuia2016}} &$23.125$& $45.165$& $45.631$& $8.371$& $0.358$& $0.618$\\
{SHFA \cite{Li2014}}&$73.322$& $\textbf{53.631}$& $\textbf{84.319}$& $\textbf{472.781}$& $\textbf{36.069}$& $\textbf{68.821}$\\
{CDLS \cite{HubertTsai2016}}&$14.492$& $3.473$& $6.269$& $12.805$& $1.010$& $2.052$\\
{ERW \cite{Kang2015}}&$12.016$& $4.476$& $4.897$& $6.481$& $1.220$& $1.494$\\
{\textbf{CDCL}}&$\textbf{83.498}$& $22.148$& $23.830$& $52.293$& $8.683$& $11.450$\\
\bottomrule[1.5pt]
\end{tabular}}
\end{table}
The experiments are conducted using Matlab R2017b in a desktop PC equipped with an Intel Core i5 CPU (at 3.1GHz) and 8GB of RAM.
Table \ref{Cost} reports the mean computational time of all methods on four cases using different numbers of source and target training samples.
Note that computational cost in \emph{Center/Univ} and \emph{Indian} cases using different numbers of training samples are reported for comparison. It is clear that \emph{CCA} and \emph{C-CCA} require the lowest time. Furthermore, among the HDA baselines, the \emph{SHFA} method requires the largest time in most cases. Since only partial unlabeled samples are used for testing in the \emph{KEMA} and \emph{SHFA} methods, to obtain a complete classification map requires much more time to obtain a complete classification map using the two methods. Compared with the \emph{ERW} method, the computational cost of the proposed \textbf{CDCL} method is 5$\sim$10 times longer. This is due to the fact that the \emph{ERW} process is applied twice in each iteration of the \textbf{CDCL} algorithm.
\section{Conclusion}
This paper has addressed the issue of semi-supervised HDA in classification of HSI. Since it is difficult (often impossible) to obtain a sufficient number of labeled samples in practical application of HSI classifcation, we assume that a limited/small number of labeled samples are available for the source and the target domains. 
The proposed method works in a  collaborative manner integrating semi-supervised learning and cross domain learning. To be specific, two main strategies are employed. Firstly, given the limited number of labeled samples in the target HSI, an intuitive strategy for obtaining more reliable pseudo-labeled samples is defined based on a comparison between RW and ERW segmentation results. Secondly, in order to obtain a better transfer ability and class separation of projected samples, C-CCA is applied for cross domain learning. The two strategies are integrated in an iterative process, in which training set and target clusters are updated until convergence. 

The experiments have been conducted on four real HSIs, i.e. Pavia University and City Center, Salinas and Indian Pines images. To explore the adaptation capacity of the proposed \textbf{CDCL} method, different numbers of training samples are randomly selected in the four cases. The proposed method achieved much better performance compared with the state-of-the-art HDA methods as well as the \emph{ERW} method.

However, the proposed method has a typical limitation of DA technique. It assumes that the labeled samples obtained by RW-based pseudolabeling are reasonably accurate, which may be a invalid assumption in some real cases. However, since the C-CCA is based on the pairwise correspondences within a cluster across domains, the source clusters can be well aligned with the corresponding target clusters even if there are few mislabelled samples in target clusters. It can be deduced that the proposed method is robust and still effective with few mislabelled samples obtained by pseudolabeling. A future development of this research is to find effective strategy to extend the proposed \textbf{CDCL} method to deep features and in pursuit of few mislabelled samples, which may result in further improvement of classification accuracy.
% or

\section*{Acknowledgment}
The authors would like to thank Prof. P. Gamba from the University of Pavia for providing the ROSIS data.


\ifCLASSOPTIONcaptionsoff
  \newpage
\fi



%\begin{thebibliography}
\bibliographystyle{IEEEtran}
\bibliography{tgrs}
%\end{thebibliography}
%\begin{thebibliography}{1}

  %0.5em minus 0.4em\relax Harlow, England: Addison-Wesley, 1999.

%\end{thebibliography}


\begin{IEEEbiography}[{\includegraphics[width=1in,height=1.25in,clip,keepaspectratio]{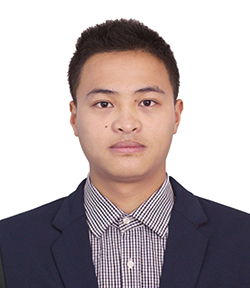}}]{Yao Qin}
%\begin{IEEEbiographynophoto}{Yao Qin}
(SM' 16) received the B.S degree in Information Engineering from Shanghai Jiaotong University, Shanghai, China, in 2013 and the M. S. degree in Information and Communication Engineering from College of Electronic Science, National University of Defense Technology (NUDT), Changsha, China, in 2015. Currently, he is pursuing the Ph. D degree and is a visiting Ph. D. with Remote Sensing Laboratory, Department of Information Engineering and Computer Science, University of Trento, Trento, Italy. His research interests include remote sensing image classification and domain adaptation.
\end{IEEEbiography}
\begin{IEEEbiography}[{\includegraphics[width=1in,height=1.25in,clip,keepaspectratio]{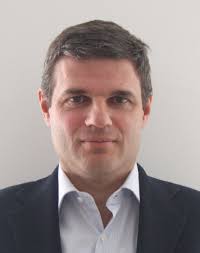}}]{Lorenzo Bruzzone}
(S'95-M'98-SM'03-F'10) received the Laurea (M.S.) degree in electronic engineering (\emph{summa cum laude}) and the Ph.D. degree in telecommunications from the University of Genoa, Italy, in 1993 and 1998, respectively. 

He is currently a Full Professor of telecommunications at the University of Trento, Italy, where he teaches remote sensing, radar, and digital communications. Dr. Bruzzone is the founder and the director of the Remote Sensing Laboratory in the Department of Information Engineering and Computer Science, University of Trento. His current research interests are in the areas of remote sensing, radar and SAR, signal processing, machine learning and pattern recognition. He promotes and supervises research on these topics within the frameworks of many national and international projects. He is the Principal Investigator of many research projects. Among the others, he is the Principal Investigator of the \emph{Radar for icy Moon exploration} (RIME) instrument in the framework of the \emph{JUpiter ICy moons Explorer} (JUICE) mission of the European Space Agency. He is the author (or coauthor) of 215 scientific publications in referred international journals (154 in IEEE journals), more than 290 papers in conference proceedings, and 21 book chapters. He is editor/co-editor of 18 books/conference proceedings and 1 scientific book. His papers are highly cited, as proven form the total number of citations (more than 25000) and the value of the h-index (74) (source: Google Scholar). He was invited as keynote speaker in more than 30 international conferences and workshops. Since 2009 he is a member of the Administrative Committee of the IEEE Geoscience and Remote Sensing Society (GRSS). 

Dr. Bruzzone ranked first place in the Student Prize Paper Competition of the 1998 IEEE International Geoscience and Remote Sensing Symposium (IGARSS), Seattle, July 1998. Since that he was recipient of many international and national honors and awards, including the recent IEEE GRSS 2015 Outstanding Service Award and the 2017 IEEE IGARSS Symposium Prize Paper Award. Dr. Bruzzone was a Guest Co-Editor of many Special Issues of international journals. He is the co-founder of the IEEE International Workshop on the Analysis of Multi-Temporal Remote-Sensing Images (MultiTemp) series and is currently a member of the Permanent Steering Committee of this series of workshops. Since 2003 he has been the Chair of the SPIE Conference on Image and Signal Processing for Remote Sensing. He has been the founder of the IEEE Geoscience and Remote Sensing Magazine for which he has been Editor-in-Chief between 2013-2017. Currently he is an Associate Editor for the IEEE Transactions on Geoscience and Remote Sensing. He has been Distinguished Speaker of the IEEE Geoscience and Remote Sensing Society between 2012-2016.
\end{IEEEbiography}
\begin{IEEEbiographynophoto}{Biao Li}
was born in Zhejiang, China, in 1968. He received the Ph.D. degrees from the
School of Electronic Science, National University of Defense
Technology, Changsha, China, in 1998, where he is currently a full professor. His current research
interests include signal processing and infrared image processing.
\end{IEEEbiographynophoto}
\begin{IEEEbiographynophoto}{Yuanxin Ye}
 (M'17) received the B.S. degree in remote sensing science and technology from Southwest Jiaotong University, Chengdu, China, in 2008, and the Ph.D. degree in photogrammetry and remote sensing from Wuhan University, Wuhan,
China, in 2013.

Since 2013, he has been an Assistant Professor with the Faculty of Geosciences and Environmental Engineering, Southwest Jiaotong University. He is currently a Post-Doctoral Fellow with the Remote Sensing Laboratory, Department of Information
Engineering and Computer Science, University of Trento, Trento, Italy. His research interests include remote sensing image processing, image registration, feature extraction, and change detection.

Dr. Ye was a recipient of the ISPRS Prizes for Best Papers by Young Authors of the 23th International Society for Photogrammetry and Remote Sensing Congress (Prague, 2016).
\end{IEEEbiographynophoto}
%\newpage


%\vfill


\end{document}